\documentclass[%
 aip,
 amsmath,amssymb,
 reprint,%
]{revtex4-1}

\usepackage{graphicx}
\usepackage{dcolumn}
\usepackage{bm}

\usepackage[utf8]{inputenc}
\usepackage[T1]{fontenc}
\usepackage{mathptmx}
\usepackage{etoolbox}

\makeatletter
\def\@email#1#2{%
 \endgroup
 \patchcmd{\titleblock@produce}
  {\frontmatter@RRAPformat}
  {\frontmatter@RRAPformat{\produce@RRAP{*#1\href{mailto:#2}{#2}}}\frontmatter@RRAPformat}
  {}{}
}%
\makeatother
\begin{document}

\preprint{AIP/123-QED}

\title[Information Evolution in Complex Networks]{Information Evolution in Complex Networks}
\author{Yang Tian}
\email{tiany20@mails.tsinghua.edu.cn}
\affiliation{ 
Department of Psychology \& Tsinghua Laboratory of Brain and Intelligence, Tsinghua University, Beijing, 100084, China.}%
\affiliation{Laboratory of Advanced Computing and Storage, Central Research Institute, 2012 Laboratories, Huawei Technologies Co. Ltd., Beijing, 100084, China.}
\author{Guoqi Li}%
 \email{guoqi.li@ia.ac.cn}
\affiliation{ 
Institute of Automation, Chinese Academy of Science, Beijing, 100190, China.}%
\affiliation{ 
University of Chinese Academy of Science, Beijing, 100049, China.}%
\author{Pei Sun}%
 \email{peisun@tsinghua.edu.cn}
\affiliation{ 
Department of Psychology \& Tsinghua Laboratory of Brain and Intelligence, Tsinghua University, Beijing, 100084, China.}%

\date{\today}

\begin{abstract}
Many biological phenomena or social events critically depend on how information evolves in complex networks. However, a general theory to characterize information evolution is yet absent. Consequently, numerous unknowns remain about the mechanisms underlying information evolution. Among these unknowns, a fundamental problem, being a seeming paradox, lies in the coexistence of local randomness, manifested as the stochastic distortion of information content during individual-individual diffusion, and global regularity, illustrated by specific non-random patterns of information content on the network scale. Here, we attempt to formalize information evolution and explain the coexistence of randomness and regularity in complex networks. Applying network dynamics and information theory, we discover that a certain amount of information, determined by the selectivity of networks to the input information, frequently survives from random distortion. Other information will inevitably experience distortion or dissipation, whose speeds are shaped by the diversity of information selectivity in networks. The discovered laws exist irrespective of noise, but the noise accounts for the intensification. We further demonstrate the ubiquity of our discovered laws by analyzing the emergence of neural tuning properties in the primary visual and medial temporal cortices of animal brains and the emergence of extreme opinions in social networks.
\end{abstract}

\maketitle

\begin{quotation}
Information dynamically evolves during its diffusion in complex networks (e.g., from individuals to individuals). The evolution process of information content creates various biological (e.g., spatial heterogeneity of neural tuning properties in the brain) and social phenomena (e.g., opinion polarization in social networks), shaping complex systems on multiple scales. However, underlying mechanisms of information evolution are little known because scientists lack a general theory to characterize information evolution yet. To lay a foundation for future explorations, we develop an information evolution characterization framework based on non-linear network dynamics theory and information theory. This framework enables us to quantitatively analyze information evolution dynamics and offers a possible explanation of the coexistence of randomness and regularity of information evolution in complex networks. Applying this framework, we suggest that representative biological and social phenomena in distinct complex systems may be uniformly explained by several discovered laws of information evolution.
\end{quotation}

\section{Introduction}\label{SEC1}
Information diffusion, as the name suggests, describes the spreading of information (e.g., viral memes or rumors) among individuals in complex networks \cite{zhang2016dynamics}. Challenge topics in various disciplines, such as communications, collective actions, and public sentiments, can be abstracted as information diffusion \cite{granovetter1978threshold,rogers2010diffusion,bakshy2012role,daley1964epidemics,backstrom2006group}. Rooted in physics \cite{goffman1964generalization,pastor2015epidemic,pastor2001epidemic,albert2002statistical}, the study of information diffusion has seen fruitful applications in biological and social sciences as well \cite{goldenberg2001talk,saito2008prediction,richardson2002mining,leskovec2007dynamics,okubo2013diffusion}. 

Till now, tremendous achievements have been accomplished in the field of the diffusion dynamics of information \cite{zhang2016dynamics,guille2013information}. Important factors to shape information diffusion properties, such as the diversity of agent behaviours \cite{iribarren2009impact}, network topology \cite{karsai2011small}, information burst patterns \cite{vazquez2007impact,karsai2011small}, and individual-individual interaction characteristics \cite{wu2011says,zhou2020realistic}, have been discovered by many empirical studies. These findings have potential insights on developing models of information diffusion \cite{zhang2016dynamics}. Static models, such as independent cascade models (assume that information diffuses via specific cascades \cite{watts2002simple,krapivsky2011reinforcement}), threshold models (focus on the roles of adjacent individuals during information diffusion \cite{watts2002simple,krapivsky2011reinforcement}), epidemic-like models (treat information diffusion like epidemic spreading \cite{trpevski2010model}), are proposed for information diffusion on static networks \cite{zhang2016dynamics}. Meanwhile, adaptive models of information diffusion processes (e.g., voter-like and social segregation models \cite{gross2008adaptive,nardini2008s}) have been developed to adopt to dynamic networks \cite{zhang2016dynamics}. These works further inspire numerous complex network theories on controlling information diffusion by network manipulation approaches (e.g., control the tipping point of diffusion \cite{chakrabarti2008epidemic} or constrain diffusion processes by node immunization \cite{prakash2013fractional} and edge rewiring \cite{tong2012gelling}). Based on these theories, computational techniques have also been developed to recover network structures from empirical information diffusion data \cite{shen2014reconstructing} and identify information sources or influential information spreaders \cite{pinto2012locating,shen2016locating,guille2013sondy,de2010does}. 

In general, the studies mentioned above primarily analyze the diffusion dynamics of information entirety (e.g., if a piece of information can traverse the entire network) \cite{zhang2016dynamics}. In other words, these works treat diffusive information like a single particle rather than a compound of contents. Although this particle-like perspective has seen substantial progress in applications \cite{zhang2016dynamics,guille2013information}, there remain numerous important problems unsolvable. Among these problems, a critical one is about information evolution, a phenomenon referring to the dynamic variation of information content during its diffusion (e.g., spatial heterogeneity of neural information representation in the brain \cite{simoncelli1998model,simoncelli1996testing,simoncelli1994velocity,rust2006mt} and opinion polarization in social networks \cite{ramos2015does,hegselmann2002opinion,sirbu2013cohesion,acemouglu2013opinion,galam2008sociophysics,sznajd2000opinion}). Even though the significant roles of information content and its evolution in shaping diffusion processes have been empirically discovered \cite{centola2010spread,de2010does,melumad2021dynamics,zhang2016creates,lee1997information,nyhof2001spreading,bebbington2017sky,stubbersfield2015serial}, information evolution has not been considered in conventional information diffusion models because the particle-like information entirety is not applicable to represent content variation dynamics. Although recent efforts have been devoted to introduce information content into the analysis of information diffusion (e.g., see Refs. \cite{lagnier2018user,lagnier2013predicting,jafari2018game,wang2017social}), there is few theoretical exploration of how non-constant information contents evolve during diffusion.

Consequently, much unknown remains about the mechanisms underlying information evolution. A fundamental problem lies in the seeming paradox of information evolution: some regular global patterns of information evolution can naturally emerge from and robustly coexist with the random distortion of information during local individual-individual diffusion (e.g., the random distortion studied by Shannon \cite{shannon1956zero}), irrespective of whether individuals tend to create such regularity or not. It is elusive how local random distortions ultimately lead to global regularity rather than accumulate to utter disorder, especially when individuals' tendency or strategy is absent. Such spontaneous global regularities, frequently observed when information diffuses in real complex networks (e.g., the hierarchical representation of visual information in the brain \cite{ishai1999distributed,van1983hierarchical}), may originate from specific undiscovered laws governing information evolution. 

The present research pursues to serve as a starting point of analyzing information evolution and may contribute to the field in three aspects: 
\begin{itemize}
    \item[\textbf{(I)} ] To characterize the widespread information evolution phenomena in complex systems, we attempt to develop a novel framework of information evolution based on a combination of network dynamics and information theory in Sec. \ref{SEC2}. We anticipate our framework to be generally applicable to characterize information evolution in any information diffusion processes;
    \item[\textbf{(II)} ] To lay a foundation of understanding potential physics laws, we primarily analyze the emergence of global regularity from local random distortions during information evolution in Sec. \ref{SEC3}. We demonstrate that specific amount of information is invariant or frequently survives from random distortion and gradually occupy most parts of the information content to create regularity of information evolution. Other information will be inevitably distorted or dissipated during diffusion. We further reveal how information selectivity properties of a complex network intrinsically determine the amount of information invariants as well as the speeds of information distortion and dissipation. These discovered laws exist irrespective of noise, but the noise accounts for their intensification; 
    \item[\textbf{(III)} ] To demonstrate the fundamental role of the discovered laws in shaping various biological and social phenomena, we generalize our theory into representative complex networks (e.g., brains and social networks) in Sec. \ref{SEC4}. Our analysis suggests that the origin of neural tuning properties in the V1 and MT cortices of the animal brain may be a natural result of information evolution. Moreover, we show that information evolution characteristics alone are sufficient to reproduce the opinion polarization process among agents, providing a potential explanation for the emergence of extreme views in social systems.
\end{itemize}

\begin{figure*}[!t]
\centering  
\includegraphics[width=2\columnwidth]{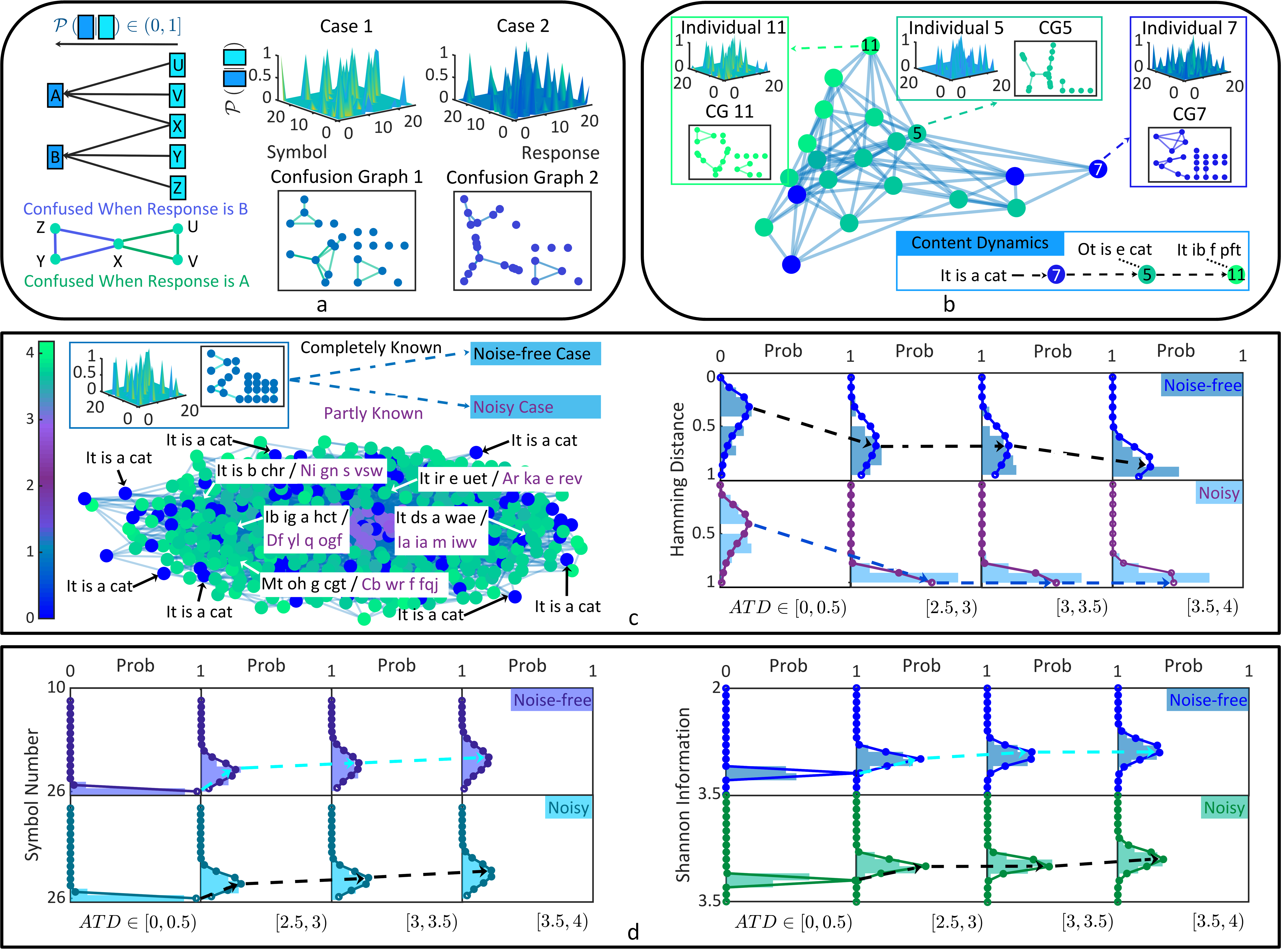}
\caption{\textbf{Information evolution during diffusion.} \textbf{a}, We illustrate an example where the confusion relations in Latin letters $\{ U,V,X \}$ or $\{ X,Y,Z \}$ occur when the response is Latin letter $A$ or $B$, respectively (left). We also show two instances of confusion graphs (middle and right). \textbf{b}, The information content evolves from the factual state ``It is a cat" to other distorted states in a diffusion path $7\rightarrow 5\rightarrow 11$. \textbf{c}, In a network of 500 individuals (color represents the average time delay (ATD) of receiving information), we set two kinds of diffusion: noise-free case (the black words in white boxes) and noisy case (the purple words in white boxes). One can find more significant information distortion in the noisy case (left). The Hamming distance between diffused information and factual information quantifies the information content variation, whose probability distribution (sampled from all individuals) gradually approaches $1$ as ATD increases. Moreover, this approaching process is sharper in the noisy diffusion than the noise-free one (right). \textbf{d}, We set a piece of new information for the complex network in \textbf{Fig. 1c}, where 26 Latin letters are uniformly distributed. The maximum number of the letters that possibly occur in the diffused information (left) and the Shannon entropy of the letter probability distribution (right) decrease with the time delay (diffusion distance), demonstrating the information dissipation.}
\end{figure*}

\section{Information evolution during diffusion}\label{SEC2}
To understand how information content might evolve, let us consider the information diffusion in a complex network (see details in Appendix \ref{SECA}). For any information, its content is principally a symbol sequence $\left(s,\ldots,s^{\prime\prime}\right)$. Any input information in the network will trigger an information diffusion process where each individual receives information from upstream individuals and passes on its response to subsequent individuals. 

As a basic network unit, each individual has multiple possible responses that vary depending on the received information (referred to as information selectivity). Meanwhile, individuals attempt to learn about (estimate) the factual input information based on what they received. After receiving symbol $s^{\prime}$, the probability for individual $i$ to respond by $s$ is $\mathsf{P}_{i}\left(s\mid s^{\prime}\right)$. This conditional probability distribution describes the information selectivity of individual $i$ (see \textbf{Fig. 1a} for an instance where we define symbol set $\mathsf{S}$ as Latin letters $\{ U,V,X,Y,Z \}$). As suggested by Shannon \cite{shannon1956zero}, an interesting situation will occur when more than one conditional probability quantity (e.g., $\mathsf{P}_{i}\left(s\mid s^{\prime}\right)$ and $\mathsf{P}_{i}\left(s\mid s^{\prime\prime}\right)$) is non-zero given different received information $s^{\prime}$ and $s^{\prime\prime}$ (e.g., see \textbf{Fig. 1a}). In this situation, there is no bijective mapping between the received information and the corresponding response. Symbol $s^{\prime}$ may be recognized as symbol $s^{\prime\prime}$ since their responses are same. This situation is referred to as information confusion in information theory \cite{shannon1956zero}. Apart from the above probabilistic description, we can also represent information confusion utilizing the confusion graph (CG) on symbol set $\mathsf{S}$ \cite{shannon1956zero,lovasz1979shannon,ahlswede1970note}. Specifically, we treat symbols as the nodes in graph $\mathsf{G}\left(\mathsf{S}\right)$ and define an edge between two nodes only if they are confused with each other (\textbf{Fig. 1a}). More details are provided in Appendix \ref{SECB}. 
 
Given these foundations, let us return to the question on how information content evolves during the diffusion process. One can consider a case where information confusion happens on individual $i$, and then the response of individual $i$ (referred to as representational information) is passed on to individual $j$. To get knowledge of the factual information received by individual $i$, individual $j$ needs to decode the representational information from individual $i$. When information confusion happens on individual $i$, the dilemma faced by individual $j$ is that one representational information corresponds to more than one possibility of factual information (\textbf{Fig. 1a}). This one-to-many mapping may make individual $j$ misestimate the factual information. Therefore, the factual information in individuals' knowledge may change when confusion happens (see Appendix \ref{SECC}). In \textbf{Fig. 1b}, we show an example of information evolution of an English sentence. One can see how the information content is gradually distorted from the factual state ``It is a cat" during diffusion. We refer to this phenomenon as information distortion. In \textbf{Fig. 1c}, we implement a more realistic and larger experiment. We consider two kinds of information diffusion. The first one requires the information selectivity of every individual to be completely known (noise-free case, which represents the ideal diffusion), while the second one does not (i.e., noisy case, which corresponds to the information diffusion in more realistic situations). One can realize the similarity between these settings and the complete/incomplete information conditions. During the diffusion, we quantify the content variation utilizing the Hamming distance \cite{hamming1950error}. We observe a continuous increase of information content changes along the diffusion pathways, which is independent of noise. However, the existence of noise accounts for accelerating information distortion (\textbf{Fig. 1c}). 

Information dissipation, a special case of distortion, is another notable phenomenon during diffusion. We define dissipation as the process where the maximum number of symbols that possibly occur in the diffused information, or the maximum information quantities possibly contained in the diffused information, decreases along the diffusion pathway (see Appendix \ref{SECD}). In \textbf{Fig. 1d}, we make all symbols in $\mathsf{S}$ (represented by Latin letters) uniformly distributed in the input information. When this information diffuses in a complex network with information confusion, information dissipation naturally emerges. One can verify this finding through symbol counting or information quantity measurement and observe its independence of noise. As expected, noise accelerate the dissipation process as well.

In summary, the above experiments demonstrate that information distortion and dissipation inevitably happen once there is information confusion during diffusion. This phenomenon occurs even during the noise-free diffusion process and is intensified if noise exists. 

\section{Invariants in information evolution}\label{SEC3}
Let us move on to the central question about the global laws governing information evolution. A possible way is to analyze what is invariant during information diffusion, irrespective of the distortion and dissipation observed above. In other words, we wonder what kind of information will not be distorted or dissipated in a given diffusion process. Because information distortion and dissipation inevitably result from information confusion, this question can be solved by measuring the maximum amount of information contents that diffuse without confusion.

Following this idea, we analyze a general case where individual $i+1$ receives information $I_{i}=\left(s_{1}^{i},\ldots,s_{l}^{i}\right)$ from individual $i$ and attempts to learn (or estimate) the factual information. The estimation necessarily requires to get knowledge of the information selectivity of individual $i$ (e.g, $\mathsf{P}_{i}\left(\cdot\mid\cdot\right)$ or $\mathsf{G}_{i}\left(\mathsf{S}\right)$). From a graphical perspective, this estimation corresponds to a subdivided reconstruction process of confusion graph $\mathsf{G}_{i}\left(\mathsf{S}\right)$. Individual $i+1$ needs to reconstruct a possible connected component where any two nodes (symbols) are confused when the response is $s_{j}^{i}$. The rest nodes excluded in this connected component are treated as isolated since response $s_{j}^{i}$ has no constraint on their confusion relations (Appendix \ref{SECD}). This reconstructed graph is referred to as $\mathsf{G}_{i}\left(\mathsf{S}\mid s_{j}^{i}\right)$, whose independence number $\alpha$ (cardinality of the largest independent set) is the maximum amount of the symbols that will not be confused given a response $s_{j}^{i}$ (\textbf{Fig. 2a}). 

\begin{figure*}[!t]
\centering  
\includegraphics[width=2\columnwidth]{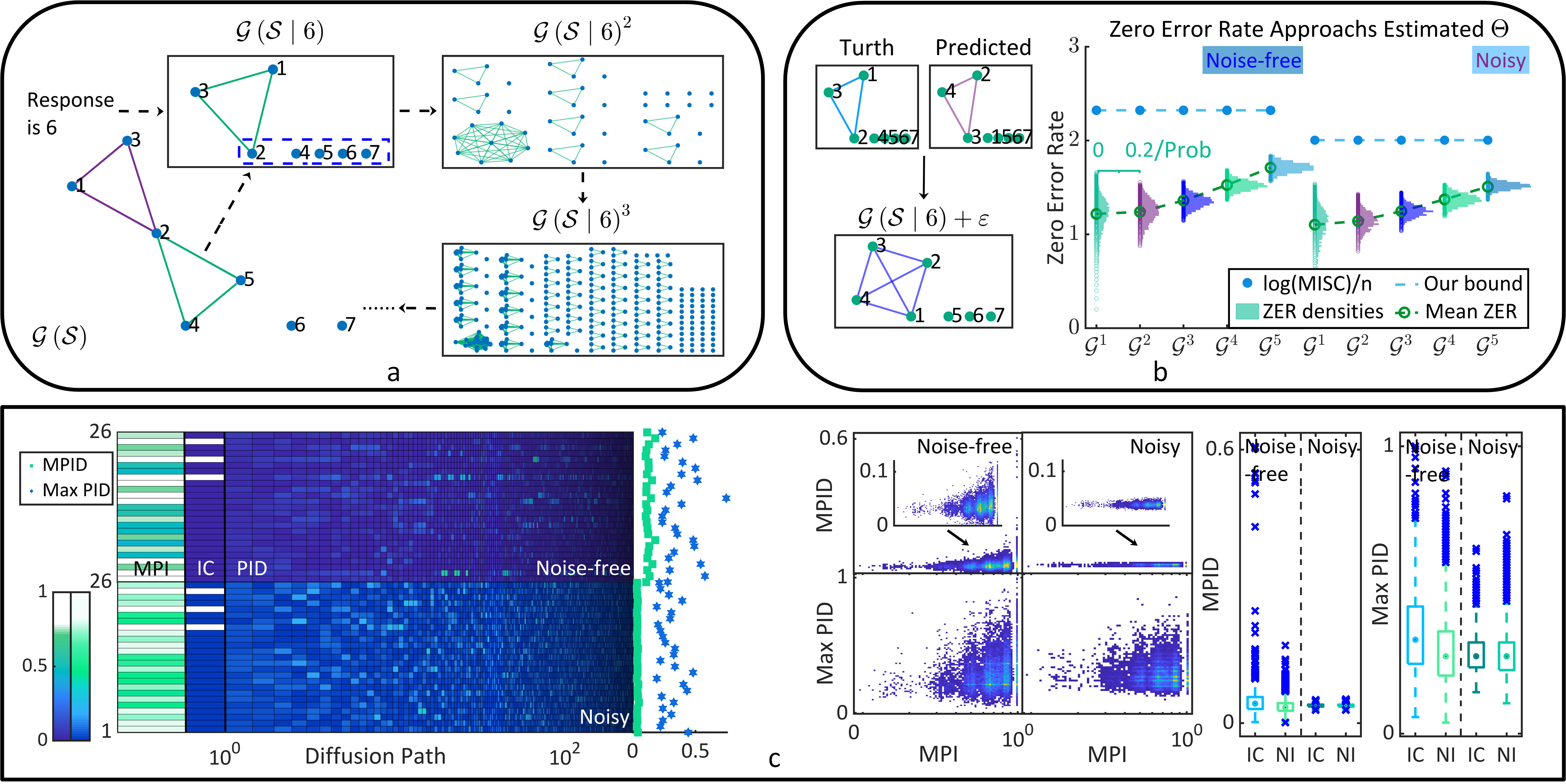}
\caption{\textbf{Invariants in information evolution.} \textbf{a}, We show the confusion graph $\mathsf{G}\left(\mathsf{S}\right)$ on a hypothetical symbol set $\mathsf{S}$ of $7$ symbols. Then we show the reconstructed confusion graph $\mathsf{G}\left(\mathsf{S}\mid 6\right)$ and its graph products. \textbf{b}, We illustrate the generation of $\mathsf{G}\left(\mathsf{S}\right)+\varepsilon$ given the ground truth graph and the predicted graph. Then, we demonstrate the validity and optimality of the bound in (\ref{EQ2}) by comparing it with the maximum zero error rate obtained by maximum independent set searching (log(MISC)/n). After the random sampling for independent sets, one can see the approaching tendency of these random sampled zero error rates (ZER) to our predicted bound (both the densities and mean value). \textbf{c}, A piece of information (26 Latin letters are uniformly distributed) diffuses in a chain of 200 individuals 500 times (see an instance in the left part). Each time the information selectivity of individuals is randomized, determining the independent set of confusion graphs and corresponding information invariants. We show that the mean probability for letter $s$ to be an invariant for every individual (MPI) positively modifies the mean/max probability for $s$ to occur in the diffused information content (MPID/Max PID). Letters that belong to the shared invariant content set of all individuals (IC) have higher MPID/Max PID values than the letters out of the common invariant content set (NI).} 
\end{figure*}

Applying the graph product $\boxtimes$, we can generalize the confusion graph of single symbols to the confusion graph of the strings of length $n$ \cite{shannon1956zero}. We implement the graph product $n-1$ times to obtain $\mathsf{G}_{i}\left(\mathsf{S}\mid s_{j}^{i}\right)^{n}=\mathsf{G}_{i}\left(\mathsf{S}\mid s_{j}^{i}\right)\boxtimes\ldots\boxtimes\mathsf{G}_{i}\left(\mathsf{S}\mid s_{j}^{i}\right)$ (\textbf{Fig. 2a}). This generalization helps formulate the maximum amount of information (no matter it is a symbol or a string) that can diffuse from individual $i$ to individual $i+1$ without confusion (zero error) given a response $s_{j}^{i}$
\begin{align}
\Theta\left(i\rightarrow i+1\mid s_{j}^{i}\right)=\sup_{n\in \mathbb{N}}\log{\sqrt[n]{\alpha\left[\left(\mathsf{G}_{i}\left(\mathsf{S}\mid s_{j}^{i}\right)+\varepsilon\right)^{n}\right]}}, \label{EQ1}
\end{align}
where $\varepsilon$ measures the noise and vanishes in the noise-free case. The reconstruct graph $\mathsf{G}_{i}\left(\mathsf{S}\mid s_{j}^{i}\right)+\varepsilon$ (see details in \textbf{Fig. 2b} and Appendix \ref{SECE}) inherits the topology of the ground truth $\mathsf{G}_{i}\left(\mathsf{S}\mid s_{j}^{i}\right)$ (since it indeed exists and shapes information diffusion) and the prediction of $\mathsf{G}_{i}\left(\mathsf{S}\mid s_{j}^{i}\right)$ by individual $i+1$ (since it affects the action pattern of individual $i+1$). Apart from that, one can also learn the basic form of equation (\ref{EQ1}) in Shannon's work \cite{shannon1956zero}.

A critical challenge concerning equation (\ref{EQ1}) is how to calculate $\Theta$. On the one hand, Shannon has suggested the difficulty of developing an analytical calculation \cite{shannon1956zero}. On the other hand, obstacles emerge inevitably in computational attempts because $\alpha$ is $NP$-hard to compute \cite{lewis1983computers}. To overcome this challenge, we turn to estimating the upper bound of $\Theta$. Because $\alpha$ is bound by the maximum clique value $\lambda$ of the graph following $\alpha\leq \lambda^{-1}$ \cite{shannon1956zero}, we can know $\log{\sqrt[n]{\alpha\left(\cdot\right)}}\leq \log\lambda^{-1}\left(\cdot\right)$. Moreover, we know that every connected component of $\left(\mathsf{G}_{i}\left(\mathsf{S}\mid s_{j}^{i}\right)+\varepsilon\right)^{n}$ ($n\in\mathbb{N}$) is a clique (since the confusion relation given a response is an equivalence relation and has transitivity, see \textbf{Fig. 2a-b}). These two properties inspire us to derive a new bound
\begin{align}
\Theta\leq\log\left[\frac{\theta+\theta\sum_{s}\left(1-\tau_{s}\right)}{\mu}\right]^{\omega}\vert \mathsf{S}\vert^{\left(1-\omega\right)}, \label{EQ2}
\end{align}
where notion $\tau_{s}=u\left[\operatorname{deg}\left(s\right)\right]$ traverses all nodes in graph $\mathsf{G}_{i}\left(\mathsf{S}\mid s_{j}^{i}\right)+\varepsilon$ (note that here $u\left(\cdot\right)$ is the unit step function). And we mark that $\omega=u\left(\sum_{s}\tau_{s}\right)$. Moreover, we pick one node that has minimum degree in the graph. Then $\mu$ measures the number of the cliques that contain this node and $\theta$ counts the cliques in the same connected component with this node (Please see Appendix \ref{SECF} for detailed derivations). Although the proposed bound in equation (\ref{EQ2}) has a complicated expression, its meaning is rather simple. In general, equation (\ref{EQ2}) suggests that the maximum number of information invariants during the diffusion from individual $i$ to individual $i+1$ is intrinsically determined by the topological properties of graph $\mathsf{G}_{i}\left(\mathsf{S}\mid s_{j}^{i}\right)+\varepsilon$, a graphical description of information selectivity characteristics of individual $i$. 

In \textbf{Fig. 2b}, the proposed bound is computationally validated by being compared with the maximum zero error (non-confusion) information rate obtained by maximum independent set searching (the Bron–Kerbosch algorithm \cite{bron1973algorithm,akkoyunlu1973enumeration}). It is easy to verify the correspondence between these two results. Furthermore, our bound is mathematically proven as a supremum when the confusion graph is not complete. In the opposite case, $\Theta=0$ can be directly obtained, making the upper bound estimation unnecessary (see Appendix \ref{SECE}). Apart from the verification, we also implement random sampling for independent sets in each graph $10^{5}$ times, calculating zero error rate samples. Consistent with Shannon's prediction, these samples approach to the upper bound as the graph product order increases.

To this point, we can conclude that information invariants are the information quantities that have not exceed the upper bound of $\Theta$. Noise can intensify distortion and dissipation as the bound in $\mathsf{G}_{i}\left(\mathsf{S}\mid s_{j}^{i}\right)+\varepsilon$ is no more than that in $\mathsf{G}_{i}\left(\mathsf{S}\mid s_{j}^{i}\right)$ (Appendix \ref{SECE}). In \textbf{Fig. 2c}, we show that information invariants are more likely to be maintained during the diffusion process, yet noise can disrupt the process.  

Furthermore, inspired by the fact that information confusion is determined by the information selectivity of individuals, we hypothesize that the diversity of information selectivity in the network plays a pivotal role in shaping information distortion, dissipation, and the convergence to invariants. The results shown in \textbf{Fig. 3} confirm our hypothesis. One can find more significant information distortion and dissipation when the information selectivity is more monotonous (\textbf{Fig. 3a}). Information invariants will occupy most parts of the diffused information content, implying a large ratio between the proportions of invariant contents (high proportion) and other contents (low proportion) in the diffused information (\textbf{Fig. 3b}). The increasing diversity of information selectivity will mitigate this process. Meanwhile, the effects of diversity will vanish due to the disturbance of noise (\textbf{Fig. 3}).

In sum, our results suggest that specific information invariants frequently survive from random distortion during diffusion. These invariants create regularity on a network scale as they gradually dominate the information content, coexisting with the random distortion during individual-individual diffusion. As for the information contents that are not invariants, their distortion and dissipation are inevitable and shaped by the diversity of information selectivity in networks.

\begin{figure*}[!t]
\centering  
\includegraphics[width=2\columnwidth]{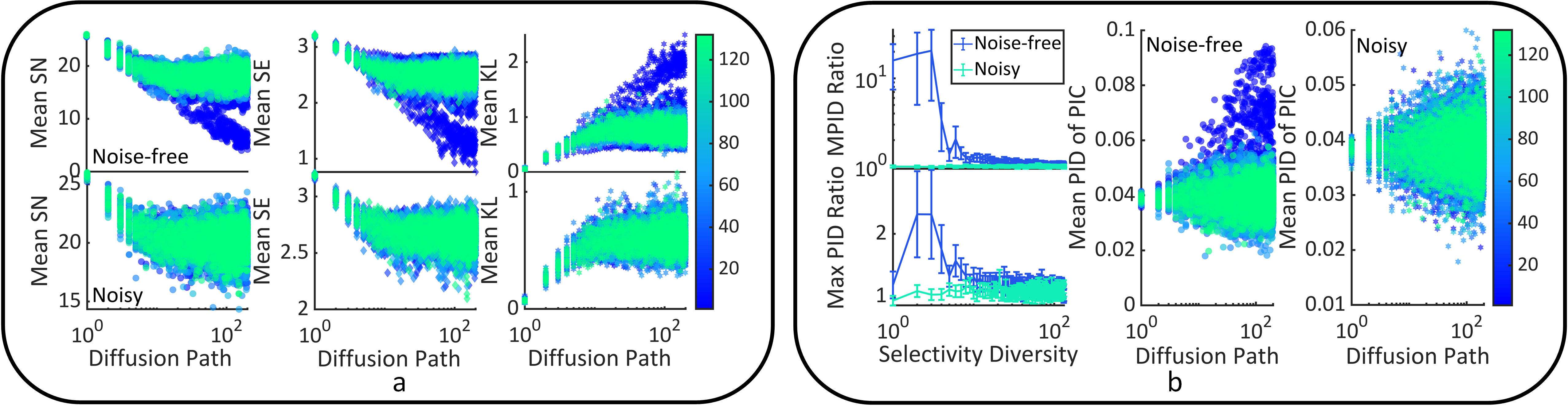}
\caption{\textbf{Information selectivity diversity shapes information evolution and the convergence to invariants.} \textbf{a}, We repeat the experiment in \textbf{Fig. 2c} 1000 times. Each time there exist $n$ kinds of information selectivity in the network ($n\in\left[1,140\right]$ is randomized). We visualize the information distortion (KL divergence between the original and the diffused symbol distributions) and dissipation (symbol number ``SN" and Shannon entropy ``SE") along the diffusion pathway (the colors of data points scale according to $n$). \textbf{b}, We generalize the scope of IC and define the potential information invariants (PIC) as the letters whose MPI values are above average. After measuring the ratios of MPID/Max PID between the letters in PIC and the letters out of PIC, we analyze these ratios as the functions of selectivity diversity $n$ (left). Note that error bars denote a quarter of the standard deviations. Moreover, we calculate the probability for a symbol to occur in the diffused information content (PID) and analyze the mean PID of the symbols in PIC during diffusion (middle and right).}  
\end{figure*}

\section{Information evolution in biological and social systems}\label{SEC4}
The above laws are discovered through an abstract information diffusion model. Below we turn to measure information diffusion in more realistic complex networks to build possible connections between our laws and biological or social phenomena. 

An example of biological system considered here is the neural pathway from the primary visual cortex (V1) to the middle temporal visual cortex (MT) in the animal brain. These two cortices are responsible for processing the orientation, direction and motion information of visual stimuli \cite{snowden1991response}. During the information diffusion from V1 to MT, a remarkable phenomenon is that the neural selectivity (a kind of information selectivity that governs neural activity profile) changes from the selectivity of the
velocity component orthogonal to the preferred
spatial orientation (simple and complex neurons in V1 \cite{adelson1982phenomenal}) to the selectivity of velocity entirety (MT neurons \cite{rust2006mt}). This variation accounts for the subdivided and staged neural representation of motion information \cite{rust2006mt}. Computationally, Simoncelli and Heeger simulate the above process in a layered neural cluster model \cite{simoncelli1998model,simoncelli1996testing,simoncelli1994velocity}. This model is generalized and experimentally-validated in a subsequent study \cite{rust2006mt}. Here we pursue to study how the modeled phenomenon naturally emerges from neural collective dynamics during information diffusion.

In \textbf{Fig. 4a}, we consider a tripartite neural cluster that is not strictly layered. The synaptic connections in this cluster are randomized. We characterize stimulus-triggered neural activities by a non-homogeneous stochastic neural network \cite{tian2021characteristics,tian2021bridging}. This framework can generate variable neural activities governed by both neural selectivity and network dynamics. In the characterization, we only preset the neural selectivity of simple neurons to affect (not completely control) their activity profiles. Consistent with previous experimental studies \cite{ringach2002orientation,ringach2002suppression}, the selectivity of each simple neuron is described by a triangular orientation tuning curve. There is no preset limitation for complex and MT neurons, providing an opportunity to explore how the characteristics of their activities emerge. We set a long enough stimulus sequence where each stimulus is a velocity vector (see Appendix \ref{SECG} for experiment settings). In \textbf{Fig. 4b}, we show the observed tuning curve of complex and MT neurons. Mathematically, we quantify how ``narrow" or ``broad" the neural selectivity is based on the variance of the normalized neural response rates for every stimulus. One can see the broadening of neural selectivity during the information diffusion from the V1 cortex to the MT cortex. To understand this phenomenon, we calculate the upper bound of $\Theta$ for each neuron based on (\ref{EQ2}). Meanwhile, we also measure the determinability rate variance of complex neurons and MT neurons, which principally quantifies the capacity of the diffusion paths from simple neurons to each complex or MT neuron to resist information distortion (see Appendix \ref{SECG}). Experimental results show that a neuron with higher response variance will also have a larger upper bound of $\Theta$, meaning that a neuron with a higher selectivity degree can pass on more undistorted information content. Moreover, a complex neuron or MT neuron that receives the information content with more distortion will have a lower selectivity degree because the determinability rate variance modifies the response variance positively. To conclude, these results suggest a possibility that the variation of neural selectivity from the V1 cortex to the MT cortex originates from the information distortion along the neural pathway.

\begin{figure*}[!t]
\centering  
\includegraphics[width=2\columnwidth]{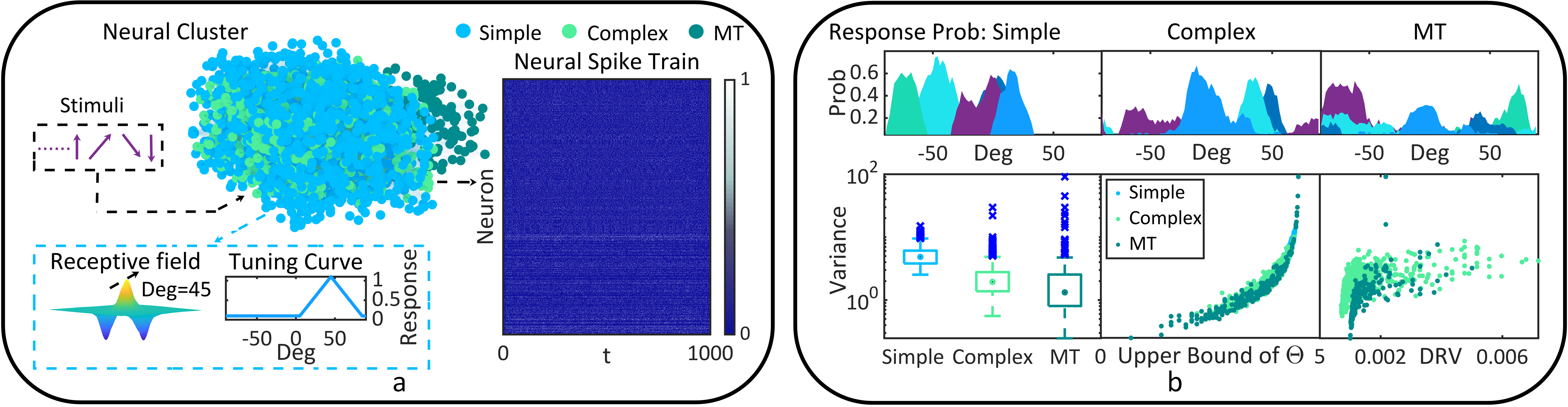}
\caption{\textbf{Information evolution in biological systems.} \textbf{a}, A neural cluster receives a random stimulus sequence and generates spikes. The receptive field and neural selectivity are merely preset for simple neurons. \textbf{b}, We respectively illustrate 5 examples of $\mathsf{P}\left(\text{Response}\mid\text{Stimulus}\right)$ for each kind of neurons. Consistent with previous studies \cite{ringach2002orientation,ringach2002suppression}, the variance of normalized neural response rates decreases from simple neurons to MT neurons. Moreover, we demonstrate that the upper bound of $\Theta$ correlates to the response variance of all neuron types positively, while the determinability rate variance (DRV) modifies the response variance of each complex or MT neuron positively.} 
\end{figure*}

\begin{figure*}[!t]
\centering  
\includegraphics[width=2\columnwidth]{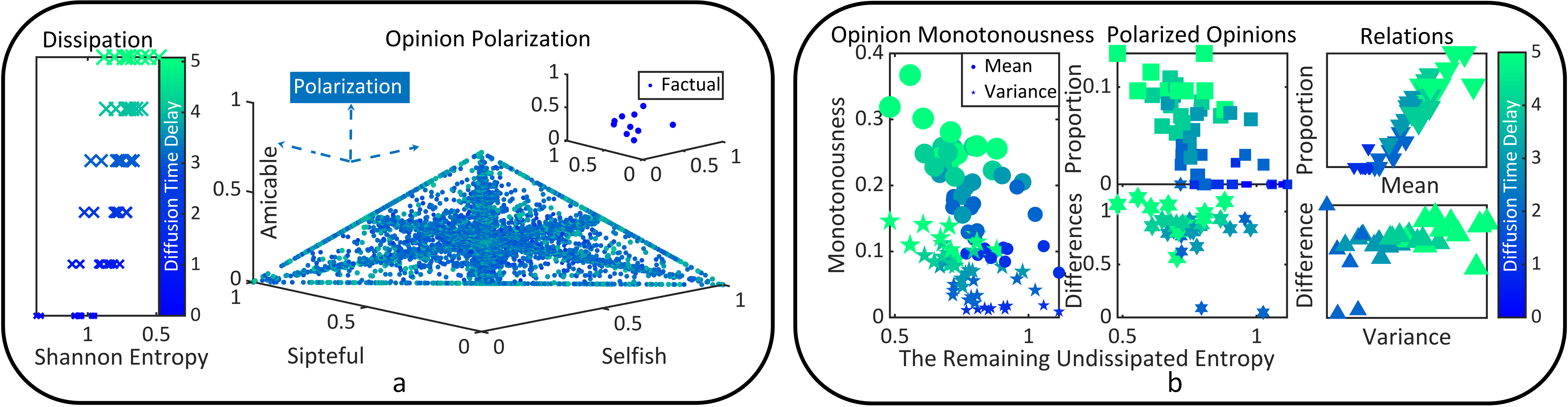}
\caption{\textbf{Information evolution in social systems.} \textbf{a}, A credit information diffusion experiment runs 10 times. Each time a random credit information begins to diffuse from several random agents. We show that the opinion polarization concerning the credit of agent $i$ happens along with information dissipation. \textbf{b}, We quantify the polarization degree utilizing the monotonousness of opinions (left). One can find that opinions become more monotonous during dissipation (see ``Mean"), and the differences between different agents' opinion monotonousness quantities gradually increase (see ``Variance"). Meanwhile, similar escalating trends can be seen in the proportion of extreme opinions among all opinions and the differences between these extreme opinions (middle). Therefore, the polarization of opinions (``Mean" and the proportion of extreme opinions) emerges during dissipation and enlarges the opinion differences (``Variance" and the differences between extreme opinions) between agents (right).}
\end{figure*}

The social phenomenon studied here is the polarization of opinions in multi-agent interactions. We implement the analysis based on the opinion concerning credit. In realistic financial, marketing, and other social activities, agents may do selfish (e.g., lie or cheat) or even spiteful (e.g., break rules for non-interest reasons) behaviors to make profits or harm others \cite{hamilton1970selfish,gardner2004spite}. These behaviors are costly \cite{fulker2021spite}, leading to a series of punishments. Among these punishments, the damage on credit and reputation is of interest for theoretical and practical reasons \cite{weigelt1988reputation,kreps1990game}. A widespread phenomenon concerning credit damage is the emergence of extreme views towards the credit of an agent. The opinions on one's credit might be polarized when credit information diffuses \cite{dandekar2013biased}. Although the agent occasionally does selfish or spiteful behaviors, its credit in others' view may still approach extraordinarily high or extremely low. This phenomenon might be caused by both psychological and physical factors \cite{dandekar2013biased,lord1979biased}. Here we explore whether information diffusion characteristics alone are sufficient to polarize the opinions on credit.

We consider a situation where agent $i$ does selfish or spiteful behaviors with probability of $p^{\prime}$ or $p^{\prime\prime}$ in a $l$-run game, respectively. Several randomly selected agents observe the game and spread the credit information. The spread information contains the credit records in every run, determining whether agent $i$ will be treated as selfish, spiteful, or amicable (see Appendix \ref{SECH} for settings). In \textbf{Fig. 5a}, we illustrate the polarization process of the opinion on the credit of agent $i$ along the diffusion pathway (each agent's opinion is the corresponding estimated factual information). Here the diffusion process is set as noisy to fit in with realistic situations. During information distortion and dissipation, the probability for agent $i$ to do selfish or spiteful behaviors in the game is driven farther from $p^{\prime}$ or $p^{\prime\prime}$. It gradually approaches $0$ or $1$, suggesting that the opinion on the credit of agent $i$ is polarized. In \textbf{Fig. 5b}, we quantify the polarization degree of opinions and attempt to measure the proportion of extreme opinions among all opinions (see Appendix \ref{SECH}). These quantitative results demonstrate that opinions are polarized along with information dissipation, during which extreme opinions naturally emerge. Meanwhile, one can see the increasing differences between different extreme opinions, suggesting the intensified disagreements between agents. In summary, the characteristics of information diffusion support reproducing the emergence of opinion polarization concerning credit and reputation. By replacing the information content with other topics, the discussed emerge process can be generalized to other kinds of opinion polarization as well.

\section{Discussion}\label{SEC5}
\subsection{Summary of our work}
The current research pursues a formal analysis of the evolution of information content when information diffuses in complex networks. Our key findings glance at the possible laws governing the dynamic evolution of information content: although information diffusion between individuals frequently involves random distortion, specific information invariants, whose quantity is bound by information selectivity characteristics in the network, are more likely to dominate the diffused information content. Any information that is not invariant will be distorted gradually, whose speed is determined by the diversity of information selectivity in the network. A particular case of such distortion that frequently occurs is dissipation, corresponding to reducing information quantity. These global laws can be observed on a network scale or along the diffusion pathway, coexisting with the stochastic distortion between individuals. Their existences do not critically rely on noise, yet noise can intensify or disturb them. 

The discovery of these potential laws begins with a formalization of the dynamics of information content utilizing the combination of network dynamics and information theory. The presented network dynamics offers a general description of the information-related interactions between individuals rather than constrained by specific information propagation models. It concentrates on the process during which each individual attempts to learn about the factual information based on the received information. Although individuals try to maintain the ground truth content while passing on the information, random distortion still originates from the confusion relations between contents \cite{shannon1956zero}. Building on Shannon's theory \cite{shannon1956zero}, we have demonstrated that the distortion and dissipation processes will inevitably emerge if the information content consists of not only information invariants. The maximum information rates of these invariants are limited by the upper bound, which has been mathematically and computationally validated as optimal. Throughout the analysis, we have distinguished these properties from the effects of noise during information diffusion, proving that the existence of information distortion and dissipation is inherently determined by the information selectivity patterns in complex networks. The speeds of information distortion and dissipation are shaped by the diversity of information selectivity in networks and might be accelerated by noises. Taken together, the theoretical framework depicted here offers a natural interpretation for the characteristics of information evolution during diffusion. 

To demonstrate the generalization capacity of our discovered laws and understand their connections with other scientific phenomena, we analyze information evolution in concrete biological and social systems. The principle that guides us through our computational experiments is to limit assumptions and explore whether complex biological or social phenomena can emerge spontaneously during information evolution. We begin by studying the origin of neural tuning properties in the V1 and MT cortices of the animal brain. Our experiment demonstrates that the neural selectivity variation process from V1 simple and complex neurons to MT neurons can be reproduced even when we only preset the tuning properties of simple neurons. For all types of neurons, their response attributes to stimuli have close relations to the upper bound of information invariants. The emergence of the neural selectivity of complex and MT neurons is a natural consequence of information distortion along the neural pathway because the response variances of these neurons are modified by the distortion situations in their cascade receptive fields. Furthermore, we turn to explore the opinion polarization and the emergence of extreme views in social systems as another instance. We contextualize our analysis under the content of credit information. We suggest that information evolution characteristics alone are sufficient to reproduce the opinion polarization process among agents. In our results, opinion polarization happens along with information dissipation. The dissipated information content enhances the monotonousness of opinions and widens the gap between different opinions. Extreme views gradually emerge in the social network as information dissipates, and one can see the increasing divergence among agents. In summary, these two computational experiments demonstrate the potential of our theory in practical applications and help us glance at the fundamental roles of information evolution in shaping biological and social systems on multiple scales.

\subsection{Related works and future directions}
Studying information evolution is a pursuit with long history and has recently received increasing attention. The first work to precisely analyze information evolution, to our best knowledge, may be Shannon's theory on errors in communication channels \cite{shannon1956zero}. At that time, information evolution is initially studied as undesired distortion detrimental to communication \cite{shannon1956zero}. Later, information evolution is extensively observed in diverse natural and artificial systems (e.g., brains \cite{ishai1999distributed,van1983hierarchical} and societies \cite{centola2010spread,de2010does,melumad2021dynamics,zhang2016creates,lee1997information}) and, therefore, begins to be studied as a natural phenomenon during information diffusion. 

As previously mentioned, one of the main challenges in the field of information evolution is the lacking of appropriate frameworks to characterize and identify potential laws underlying the dynamics of information content. Such a challenge arises from that conventional information diffusion models defined from a particle-like perspective (e.g., independent cascade \cite{watts2002simple,krapivsky2011reinforcement}, threshold \cite{watts2002simple,krapivsky2011reinforcement}, epidemic-like \cite{trpevski2010model} models and all
subsequent works based on them) are not applicable to represent information contents and their variations \cite{lagnier2018user}. Although recent models \cite{lagnier2013predicting,lagnier2018user,jafari2018game,wang2017social} begin to specify information contents during diffusion, the variation characteristics of non-constant information contents have not been explored.

In real complex networks, different information contents may experience distinct diffusion processes. The responses to a piece of information may vary across different individuals. Meanwhile, the responses of an individual may be information-content-based \cite{lagnier2018user}. These properties inspire us to develop a passing-words-like framework including information contents and information selectivity of individuals. Under this framework, one can generally understand information diffusion as a process where an individual receives information from previous individuals, generates responses according to information selectivity, attempts to learn about the factual information received by previous individuals in the meanwhile, and passes on its responses to subsequent individuals. Here information selectivity is a concept generalized from neural selectivity in neuroscience \cite{dayan2005theoretical} to characterize how the responses of an individual vary depending on received information contents. Information evolution during diffusion lies in the dynamics of the learned factual information by individuals. Similar ideas can be seen in individual-centered probabilistic models of information-content-based diffusion \cite{lagnier2018user,lagnier2013predicting}, yet there are detailed differences between these probabilistic models and our work. Specifically, these models consider information contents and information selectivity (referred to as user profiles \cite{lagnier2018user,lagnier2013predicting}) to analyze the willingness or strategies of individuals (referred to as users \cite{lagnier2018user,lagnier2013predicting}) to pass on certain information contents. Contents are defined as static elements to distinguish between different pieces of information in information space rather than variable sets of symbols to characterize potential information evolution \cite{lagnier2018user,lagnier2013predicting}. Our motivation to consider information contents as alterable symbols sets is to characterize information selectivity with probabilities for symbols to occur in responses conditional on received symbols and find potential confusion relations between symbols. Based on this definition, we can naturally relate our framework with Shannon's theory \cite{shannon1956zero}, which essentially requires symbol confusion relations to analyze errors when information diffuses from one individual to another. In sum, our first contribution compared with previous studies is to develop an applicable framework to study information evolution during diffusion based on Shannon's theory \cite{shannon1956zero}.

In the context of Shannon's theory \cite{shannon1956zero}, we can subdivide information contents as invariant and variable parts depending on information selectivity characteristics of information senders and receivers. These invariant parts can diffuse from a sender to a receiver without confusion (will not be distorted) \cite{shannon1956zero}. Therefore, analyzing invariant contents (e.g., measure the upper bound of their quantities) may serve as a feasible direction to study how global regularity emerges from local random distortions during information evolution. However, analytically measuring $\Theta$, the upper bound of amounts of information invariants, remains as a daunting challenge in information theory \cite{shannon1956zero,lovasz1979shannon}. For instance, the analytic calculation of $\Theta$ when confusion graph is a $5$-cycle has remained unsolved until Lov{\'a}sz's work \cite{lovasz1979shannon}. Meanwhile, computational attempts to derive $\Theta$ will inevitably meet obstacles because the cardinality of the largest independent set in confusion graphs, a precondition of $\Theta$ calculation, is $NP$-hard to compute \cite{lewis1983computers}. To overcome these difficulties, our second contribution is to show that every connected component of the confusion graph under our passing-words-like framework must be a clique, based on which, we propose an upper bound of $\Theta$ in (\ref{EQ2}) and further prove that this upper bound is actually a supremum when confusion graph is not complete in Appendix \ref{SECF} (when confusion graph is complete, $\Theta=0$ can be directly obtained, making the upper bound estimation unnecessary). These results enable us to efficiently measure $\Theta$ on any confusion graph that may occur during information evolution. To compare (\ref{EQ2}) with related works, we validate our supremum with the maximum zero error (non-confusion) information rate obtained by maximum independent set searching (the Bron–Kerbosch algorithm \cite{bron1973algorithm,akkoyunlu1973enumeration}). Consistency can be found between our supremum and the results of these time-consuming algorithms. Meanwhile, we implement random sampling for independent sets on confusion graphs to show that the sizes of independent sets approach to our supremum as the graph product order increases. These results are consistent with Shannon's prediction \cite{shannon1956zero}.

Based on the proposed supremum of information invariants, our third contribution compared with previous studies \cite{shannon1956zero,lovasz1979shannon} is to relate information invariants with the emerged global regularity from local random distortions during information evolution. Specifically, global regularity can be manifested as a process where information invariants survive from random distortions to gradually dominate diffused information contents. Our experiments demonstrate that global regularity can emerge on a network scale or along the diffusion pathway to coexist with random distortions between individuals. This theoretical result is comprehensible since similar phenomena have been extensively observed by empirical studies. For instance, it has been found that news information evolves as it is retold by individuals (a kind of passing-words-like process where one individual retell the news to another individual) \cite{melumad2021dynamics}. Drawing on data from over 11000 individuals, researchers discover that news information experiences random and stylistic distortions where original information is gradually dominated by specific public opinions (e.g., negative opinions) \cite{melumad2021dynamics}. Other parts of information contents are distorted or dissipated depending on how individuals retell the news (information selectivity), making the retold news become increasingly extreme \cite{melumad2021dynamics}. Similar empirical instances, to name a few, can be seen in Refs. \cite{nyhof2001spreading,bebbington2017sky,stubbersfield2015serial}. Compared with these empirical studies, our fourth contribution is to demonstrate that characteristics of information selectivity (e.g., distribution and diversity) in a complex network intrinsically determine the quantities of information invariants and the distortion speeds of variable parts of information contents during information diffusion. On the one hand, the supremum in (\ref{EQ2}) is essentially measured on confusion graphs defined by information selectivity of information senders and receivers. On the other hand, the diversity of information selectivity in a complex network shapes the speeds of information distortion and the occupancy process of information invariants in diffused information contents. These properties do not critically rely on conventional factors during information evolution, such as noise, yet noise can intensify or disturb them. Based on these properties, we can quantitatively predict information evolution according to the inherent information selectivity characteristics of arbitrary complex networks. 

As shown by our computational experiments, our theory can be naturally applied to analyze information evolution phenomena underlying neural tuning property formation in animal brains \cite{simoncelli1998model,simoncelli1996testing,simoncelli1994velocity,rust2006mt} and extreme opinion emergence in social networks \cite{ramos2015does,hegselmann2002opinion,sirbu2013cohesion,acemouglu2013opinion,galam2008sociophysics,sznajd2000opinion}. We suggest that our theory is not only applicable to abstract passing-words-like process but also to concrete complex system models (e.g., the non-homogeneous stochastic neural network \cite{tian2021characteristics,tian2021bridging} for  modelling neural dynamics). Compared with existing mathematical models proposed for simulation \cite{simoncelli1998model,simoncelli1996testing,simoncelli1994velocity,rust2006mt,ramos2015does,hegselmann2002opinion,sirbu2013cohesion,acemouglu2013opinion,galam2008sociophysics,sznajd2000opinion}, our fifth contribution is to provide possible explanations of corresponding phenomena from the perspective of information evolution.

As an exploratory study, our work has multiple limitations remaining for future explorations. A valuable direction may be further investigating the effects of network topology on information evolution. Although information invariants and all related concepts are directly determined by information selectivity characteristics of complex networks rather than network topology, we suggest that network topology may be a moderator variable to affect information evolution indirectly. This is because all concepts of information evolution are either based on or related to the confusion graphs defined by the information selectivity of information senders and receivers. Network topology defines connections between individuals to determine the one-to-one correspondence between senders and receivers and, therefore, affects confusion graphs. Moreover, a significant question for future exploration is how information distortion and dissipation are reduced when individuals in the complex network optimize the estimation of the factual information content. In realistic biological and social iterations, individuals have a strong tendency to enable themselves to learn about information as efficiently as possible. Optimizing the factual information estimation approach might enable individuals to de-correlate between information content (control the confusion relations) and raise the upper bound of information invariants. Therefore, it would be meaningful to go further into the intricate effects of individuals' active behaviors on information evolution.

\begin{figure*}[!t]
\centering  
\includegraphics[width=2\columnwidth]{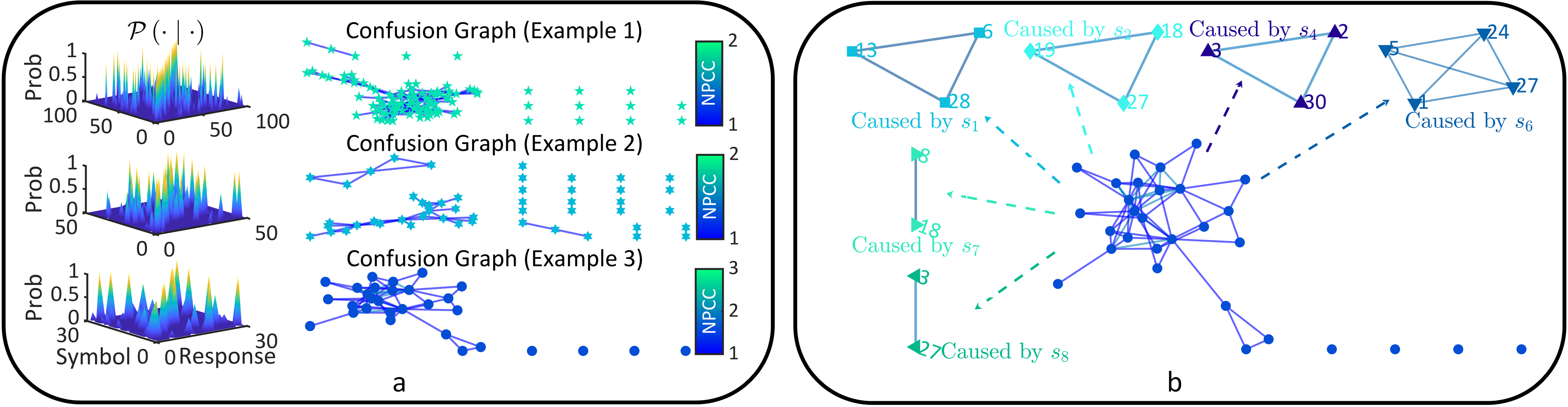}
\caption{\textbf{Information confusion.} \textbf{a}, Here we illustrate 3 examples of probability distribution $\mathsf{P}\left(\cdot\mid \cdot\right)$ on a symbol set $\mathsf{S}$ of $k$ symbols (here $k\in\{100,50,30\}$). One can see the cases where $\mathsf{P}\left(\cdot\mid \cdot\right)=1$ or $\mathsf{P}\left(\cdot\mid \cdot\right)=0$ (non-confusion) and the cases where $\mathsf{P}\left(\cdot\mid \cdot\right)\in\left(0,1\right)$ (confusion). Correspondingly, we visualize the confusion graph, where each edge scales based on the number of possible confusion cases (NPCC). For instance, $\text{NPCC}=n$ for edge $\left(s_{x},s_{y}\right)$ means that the confusion relation between symbols $s_{x}$ and $s_{y}$ might be caused by $n$ kinds of responses. \textbf{b}, When the response has been specified, one can find all the confusion relations caused by this response.}
\end{figure*}

\begin{figure*}[!t]
\centering  
\includegraphics[width=2\columnwidth]{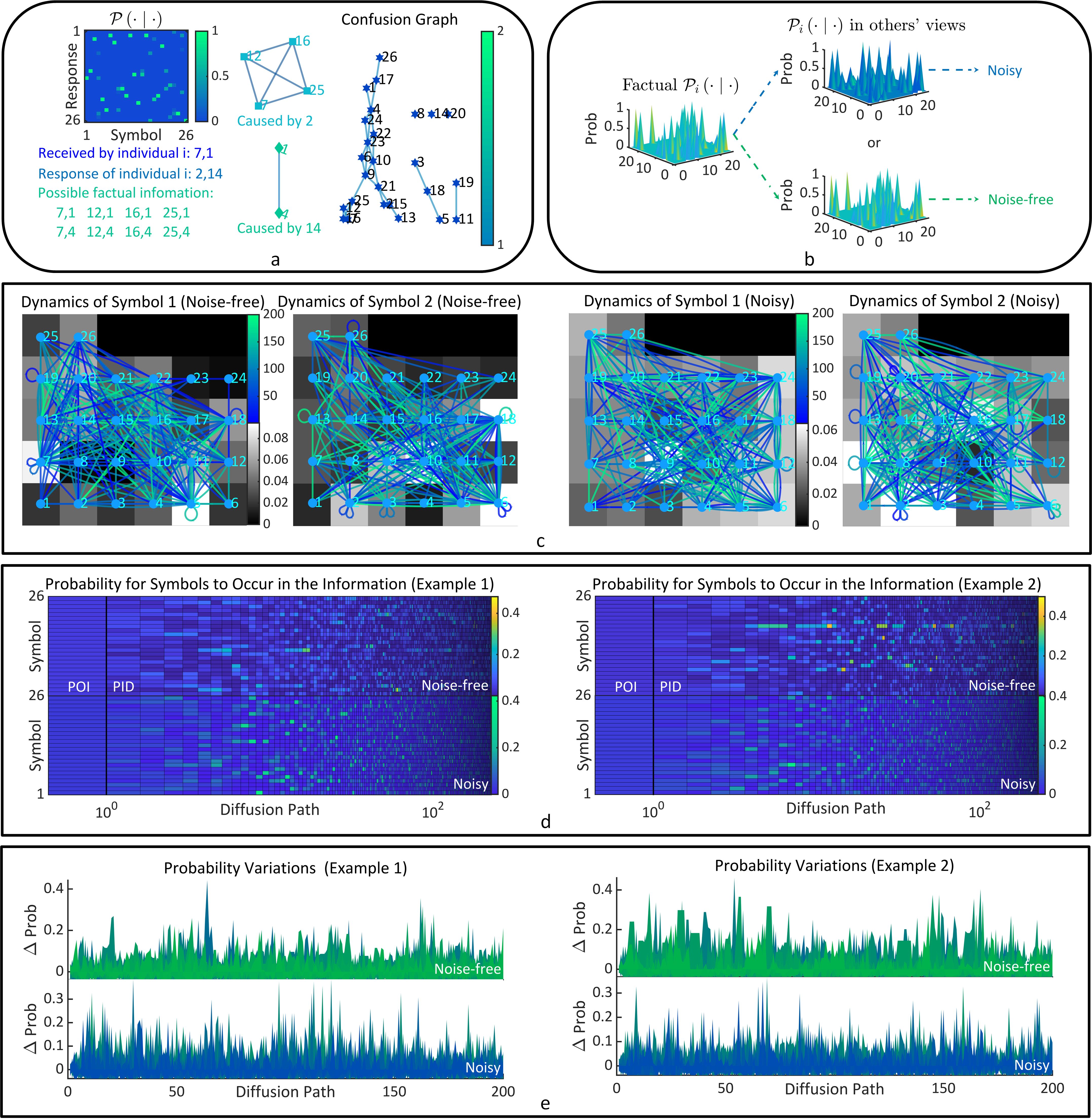}
\caption{\textbf{Information distortion.} \textbf{a}, We illustrate an example where information confusion may lead to the distortion of information content. \textbf{b}, Two kinds of information diffusion are defined in our research. The first one requires the information selectivity of each individual to be completely known (noise-free diffusion) while the second kind does not (noisy diffusion). \textbf{c}, We randomize a piece of information consisting of uniformly distributed 26 symbols and let it diffuse in a chain of 200 individuals. Then, we visualize the dynamics of two symbols in the factual information content. These symbols change along the diffusion pathway (see the digraphs in color), transforming into other symbols with specific probability (see the background matrices in gray). \textbf{d}, The probability for each symbol to occur in the information content changes from its initial quantity (POI) to other quantities during diffusion (PID). \textbf{e}, Correspondingly, we calculate the difference between PID and POI.}
\end{figure*}

\begin{acknowledgments}
Correspondence should be addressed to G.Q.L. and P.S. This project is supported by the Artificial and General Intelligence Research Program of Guo Qiang Research Institute at Tsinghua University (2020GQG1017) as well as the Tsinghua University Initiative Scientific Research Program. Authors are grateful for discussions and assistance of Drs. Yaoyuan Wang and Ziyang Zhang from the Laboratory of Advanced Computing and Storage, Central Research Institute, 2012 Laboratories, Huawei Technologies Co. Ltd., Beijing, 100084, China. 
\end{acknowledgments}

\section*{Author Declarations}
\subsection*{Conflict of Interest}
The authors have no conflicts to disclose.

\subsection*{Data Availability Statement}
The data that support the findings of this study are available from the corresponding author upon reasonable request.

\appendix
\section{Complex networks and individuals}\label{SECA}
All complex networks in our research are defined as connected random graphs utilizing the standard approach introduced by Erd{\H{o}}s and R{\'e}nyi \cite{erdHos1960evolution,cameron1997random}. For convenience, the average degree of each random network is randomized as $\big\langle\operatorname{deg}\big\rangle\in\left[5,20\right]$. In most cases, there is no restriction on the local topology characteristics of a random network during initialization (the only exception is the neural cluster used in our neuroscience experiment, where we design the network topology with realistic neural settings). 

Information diffusion starts from a set of randomly selected individuals and gradually traverses all individuals in the network. For individual $j$ who receives information indirectly, we quantify the time delay of information diffusion as $\eta\left(j\right)=\big\langle\varphi\left[\mathsf{W}\left(i\rightarrow j\right)\right]\big\rangle_{i}$, where $\mathsf{W}\left(i\rightarrow j\right)$ is the shortest path from $i$ to $j$, notion $\langle\cdot\rangle_{i}$ measures the expectation value by traversing every individual $i$ that receives factual information directly, and $\varphi\left(\cdot\right)$ denotes the time delay weighting mapping. Our research uses a simplified definition $\varphi\left[\mathsf{W}\left(i\rightarrow j\right)\right]=\vert\mathsf{W}\left(i\rightarrow j\right)\vert$, meaning that the information diffusion between two neighbors costs a duration of $1$ and the time cost on a path equals the path length.

Every individual in the complex network has its information selectivity, which can be principally described by
$\mathsf{P}\left(s_{x}\mid s_{y}\right)$. Information selectivity is free to be applied in modeling preference, strategy, and other individualities. Except for the neuroscience experiment where we design the randomization of $\mathsf{P}\left(s_{x}\mid s_{y}\right)$ with neural characteristics, the randomization in common cases is implemented without any restriction. The information selectivity properties of any two individuals in a complex network are not necessarily same. Therefore, the network can be either homogeneous or heterogeneous.

The process for an individual to estimate the factual information based on the received information is defined by a naive Bayesian inference, namely $\mathsf{P}\left(s_{y}\mid s_{x}\right)\propto\mathsf{P}\left(s_{x}\mid s_{y}\right)\mathsf{P}\left(s_{y}\right)\mathsf{P}\left(s_{x}\right)^{-1}$. The estimated factual information by individual $j$ represents the beliefs of $j$ on the ground truth information content. 

In summary, information diffusion refers to a process where each individual receives information and passes on its response following information selectivity. Meanwhile, this individual attempts to learn about the factual information received by previous individuals. Please note that the passed-on response can be either same as (e.g., when this individual transmits what it knows) or different from (e.g., when this individual does not share its knowledge directly) the learned factual information. This property of practical significance can be implemented based on information selectivity. While analyzing information content variations during diffusion, what we measure is the dynamics of the learned factual information by individuals.

 \section{Probabilistic and graphical descriptions of information confusion}\label{SECB}
 
 In the main text, we have sketched confusion relations from probabilistic and graphical perspectives. Here we elaborate the detailed definitions. 

\begin{figure*}[!t]
\centering  
\includegraphics[width=2\columnwidth]{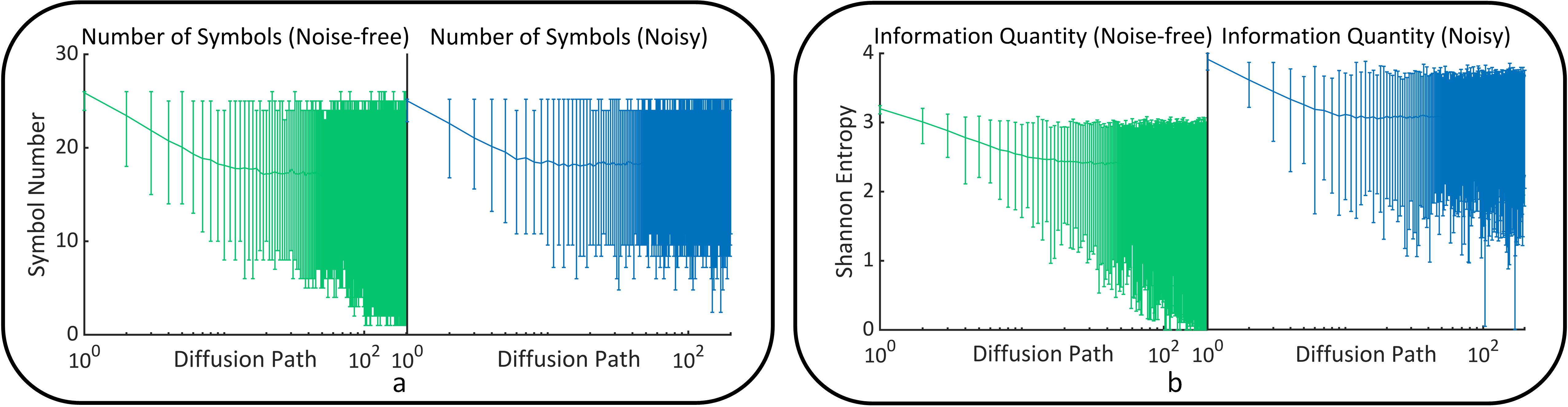}
\caption{\textbf{Information dissipation.} \textbf{a}, The results of symbol counting indicate the decreasing process of the diversity of symbols in information content during diffusion. \textbf{b}, Correspondingly, one can see that the information quantity contained in information content decreases as well.}
\end{figure*}

Let us consider a finite symbol set $\mathsf{S}$. We define that two symbols $s_{x},\;s_{y}\in\mathsf{S}$ are confused with each other if they may lead to the same response $s_{z}\in\mathsf{S}$ of a certain individual $i$. From the probabilistic aspect, this condition requires $\mathsf{P}_{i}\left(s_{z}\mid s_{x}\right)\in\left(0,1\right]$ and $\mathsf{P}_{i}\left(s_{z}\mid s_{y}\right)\in\left(0,1\right]$. After receiving response $s_{z}$ from individual $i$ (this means $\mathsf{P}\left(s_{z}\right)=1$, implying $\mathsf{P}_{i}\left(s_{x}\mid s_{z}\right)\in\left(0,1\right)$ and $\mathsf{P}_{i}\left(s_{y}\mid s_{z}\right)\in\left(0,1\right)$), individual $j$ can not confirm which symbol causes this response if there is no other information. Therefore, there is no confusion relation between symbols $s_{x}$ and $s_{y}$ when the response is $s_{z}$ if 
\begin{align}
    \{\mathsf{P}\left(s_{x}\mid s_{z}\right),\mathsf{P}\left(s_{y}\mid s_{z}\right)\}\cap\{0,1\}\neq\emptyset. \label{AEQ1}
\end{align}

Let us assume that individual $j$ attempts to estimate what individual $i$ receives, then symbol $s_{x}$ may be recognized as symbol $s_{y}$. This misrecognition between symbols, or referred to as information confusion, is studied by Shannon from the graphical perspective \cite{shannon1956zero}. Specifically, Shannon represents symbols by the nodes in graph $\mathsf{G}\left(\mathsf{S}\right)$ and defines an edge between two nodes if there is a confusion relation. Our research builds on Shannon's work and further distinguish between different types of confusion relations based on their corresponding causes (see \textbf{Fig. 6}).

\section{Information distortion}\label{SECC}
Let us assume that there exist information confusion phenomena caused by the information selectivity of an individual. From other individuals' perspectives, information confusion creates multiple possibilities when they attempt to learn about the factual information received by the previous individual (see \textbf{Fig. 7a}). Without other auxiliary clues, these individuals might misestimate the factual information, leading to information content variation. This phenomenon is referred to as information distortion in our research. Please note that the noise during information transmission is excluded in the above analysis, implying that the existence of information distortion is independent of noise. Taking noise into consideration, we can further distinguish between noise-free and noisy information diffusion and compare information distortion during these two kinds of processes (see \textbf{Fig. 7b}).

Here we show several distortion processes of the symbols in information content, where one can see the transformation of the original symbols to other symbols due to misestimation (see \textbf{Fig. 7c}). Meanwhile, we also illustrate the dynamics of the probability for each symbol to occur in the information, indicating the effects of information distortion on information content during diffusion (see \textbf{Fig. 7d-e}).

\section{Information dissipation}\label{SECD}
In our research, we define information dissipation as a kind of distortion process where the maximum number of symbols that possibly occur in the diffused information, or the maximum information quantities possibly contained in the diffused information, gradually decreases. 

During information diffusion, the original probability distribution of symbols in the information content experiences complex changes (e.g., see \textbf{Fig. 7d}). This variation process usually behaves as decreasing. In \textbf{Fig. 8}, we implement symbol counting and information quantity measurement during the experiment in \textbf{Fig. 7} to demonstrate the existence of information dissipation.

\section{Confusion graph reconstruction during estimation}\label{SECE}
As described in our main text and Appendix \ref{SECA}, the process for other individuals to learn (or estimate) the factual information received by individual $i$ based on its response $I_{i}=\left(s_{1}^{i},\ldots,s_{l}^{i}\right)$ virtually requires getting knowledge of the information selectivity of individual $i$ (e.g, $\mathsf{P}_{i}\left(\cdot\mid\cdot\right)$ or $\mathsf{G}_{i}\left(\mathsf{S}\right)$). In a graphical perspective, this estimation is equivalent to a subdivided reconstruction of confusion graph $\mathsf{G}_{i}\left(\mathsf{S}\right)$. For each symbol $s_{j}^{i}$ in the response, the corresponding reconstructed graph $\mathsf{G}_{i}\left(\mathsf{S}\mid s_{j}^{i}\right)$ only contains the confusion relations caused by this symbol (e.g., see \textbf{Fig. 9a}).

Considering the potential noise during diffusion, we can further define $\mathsf{G}_{i}\left(\mathsf{S}\mid s_{j}^{i}\right)+\varepsilon$ to fit in with noisy information diffusion. Here $\varepsilon$ measures the noise and vanishes in the noise-free case. Our research suggests that the confusion graph $\mathsf{G}_{i}\left(\mathsf{S}\mid s_{j}^{i}\right)$ in other individuals' views may not be the same as the factual one due to noises (e.g., see \textbf{Fig. 9b}). The factual confusion graph governs the actual information diffusion process while the imaginary one affects other individuals' behaviours. Therefore, the topology information of these two confusion graphs should be both taken into consideration. Specifically, we define 
\begin{align}
&E\left[\mathsf{G}_{i}\left(\mathsf{S}\mid s^{i}_{j}\right)+\varepsilon\right]=\Big\lbrace \left(s,s^{\prime}\right)\mid\mathsf{W}\left(s\rightarrow s^{\prime}\right)\in\notag\\ &W\left[\mathsf{G}_{i}\left(\mathsf{S}\mid s^{i}_{j}\right)\right]\cup W\left[\widehat{\mathsf{G}}_{i}\left(\mathsf{S}\mid s^{i}_{j}\right)\right]\Big\rbrace, \label{AEQ2}
\end{align}
where $\widehat{\mathsf{G}}$ stands for the prediction of confusion graph $\mathsf{G}$, notion $E\left(\cdot\right)$ denotes the edge set, and $W\left(\cdot\right)$ denotes the path set (see \textbf{Fig. 9b}).

We further generalize the confusion graphs of symbols to the confusion graphs of strings following Shannon's idea \cite{shannon1956zero}. Our generalization is implemented based on the graph product $\boxtimes$ (e.g., see \textbf{Fig. 9c}). The precondition for any two tuples $\left(s_{p},s_{q}\right),\;\left(s_{b},s_{d}\right)\in\mathsf{S}\times\mathsf{S}$ to be connected in the graph product $\mathsf{G}_{i}\left(\mathsf{S}\mid s^{i}_{j}\right)\boxtimes\mathsf{G}_{i}\left(\mathsf{S}\mid s^{i}_{j}\right)$ is one of the following cases: 
\begin{itemize}
\item $s_{p}=s_{b}$ and $s_{q}$ is connected with $s_{d}$ in graph $\mathsf{G}_{i}\left(\mathsf{S}\mid s^{i}_{j}\right)$;
\item $s_{q}=s_{d}$ and $s_{p}$ is connected with $s_{b}$ in graph $\mathsf{G}_{i}\left(\mathsf{S}\mid s^{i}_{j}\right)$;
\item $s_{p}$ and $s_{q}$ are respectively connected with $s_{b}$ and $s_{d}$ in graph $\mathsf{G}_{i}\left(\mathsf{S}\mid s^{i}_{j}\right)$.
\end{itemize}
This generalization allows to define 
\begin{align}
\mathsf{G}_{i}\left(\mathsf{S}\mid s^{i}_{j}\right)^{n}=\mathsf{G}_{i}\left(\mathsf{S}\mid s^{i}_{j}\right)\boxtimes\ldots\boxtimes\mathsf{G}_{i}\left(\mathsf{S}\mid s^{i}_{j}\right), \label{EQ3}
\end{align}
representing the confusion relations between strings of length $n$ (e.g., see \textbf{Fig. 9c}).

\begin{figure*}[!t]
\centering  
\includegraphics[width=2\columnwidth]{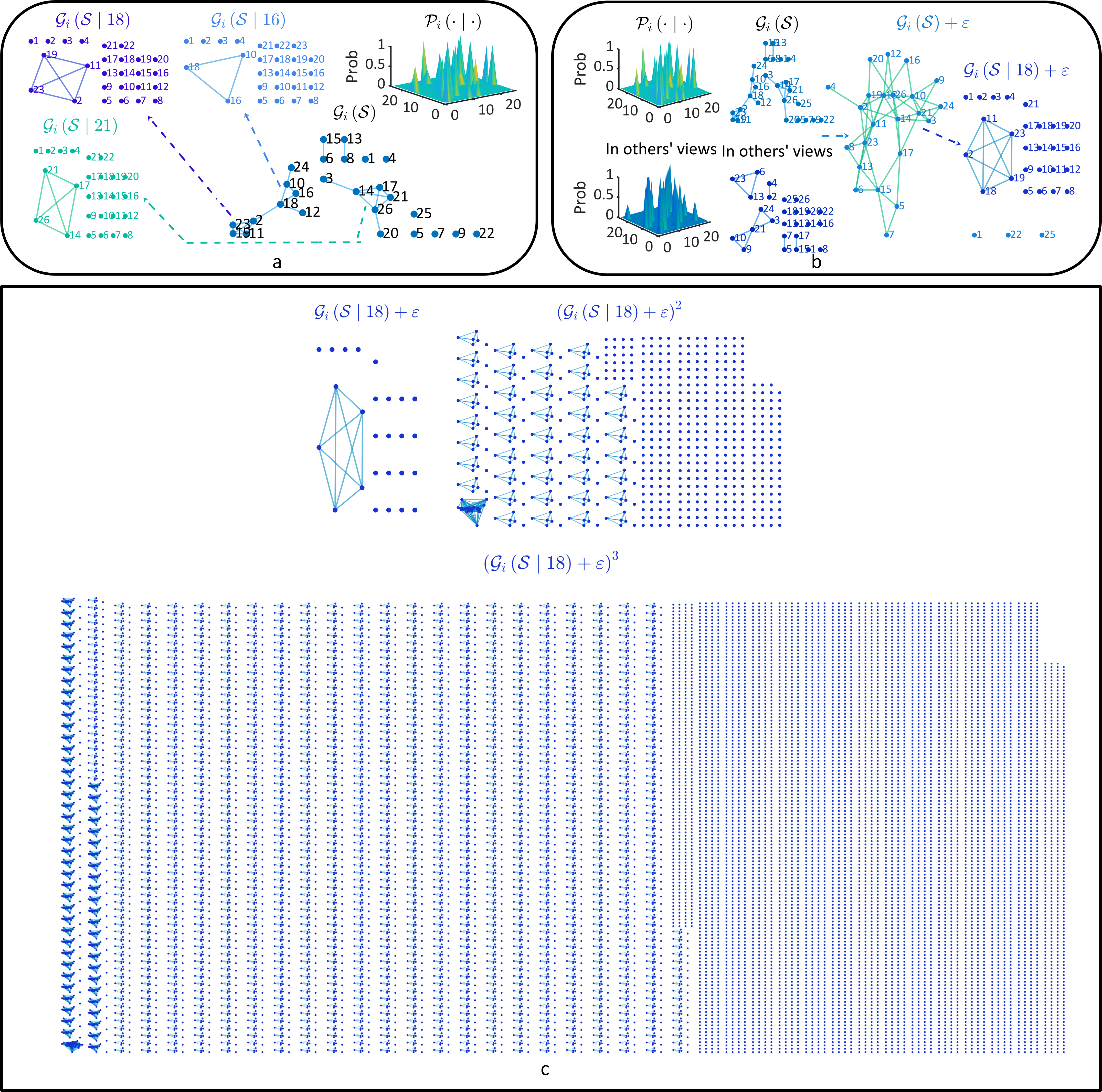}
\caption{\textbf{Information dissipation.} \textbf{a}, We illustrate several reconstructed confusion graphs $\mathsf{G}_{i}\left(\mathsf{S}\mid\cdot\right)$ caused by specific responses of individual $i$. \textbf{b}, We show how $\mathsf{G}_{i}\left(\mathsf{S}\right)+\varepsilon$ and $\mathsf{G}_{i}\left(\mathsf{S}\mid\cdot\right)+\varepsilon$ inherit topology information from the ground truth confusion graph $\mathsf{G}_{i}\left(\mathsf{S}\right)$ and the confusion graph in other individuals' views. \textbf{c}, We illustrate the graph products of a reconstructed confusion graph up to $3$-order.}
\end{figure*}

\section{Upper bound of information invariants}\label{SECF}
In our main text, we have formalized the maximum amount of information (no matter if it is a symbol or string) that can diffuse from individual $i$ to other individuals without confusion given a response $s_{j}^{i}$
\begin{align}
\Theta\left(i\rightarrow i+1\mid s_{j}^{i}\right)=\sup_{n\in \mathbb{N}}\log{\sqrt[n]{\alpha\left[\left(\mathsf{G}_{i}\left(\mathsf{S}\mid s_{j}^{i}\right)+\varepsilon\right)^{n}\right]}}, \label{EQ4}
\end{align}
where $\alpha\left(\cdot\right)$ denotes the independence number of confusion graph (cardinality of the largest independent set), measuring the maximum amount of the symbols that will not be confused given a response $s_{j}^{i}$. Equation (\ref{EQ4}) is firstly introduced in Shannon's work \cite{shannon1956zero}. This definition helps explore what will be invariant during information diffusion. 

A daunting challenge lies in that the analytic calculation of $\Theta$ is principally difficult (e.g, Shannon failed to measure $\Theta$ on $5$-cycle \cite{shannon1956zero}. This problem has remained unsolved until Lov{\'a}sz's work \cite{lovasz1979shannon}). On the other hand, any computational attempt will inevitably meet obstacles because $\alpha$ is $NP$-hard to compute \cite{lewis1983computers}. We have built on Shannon's work \cite{shannon1956zero} to estimate the upper bound of $\Theta$
\begin{align}
\Theta\left(i\rightarrow i+1\mid s_{j}^{i}\right)\leq\log\left[\frac{\theta+\theta\sum_{s}\left(1-\tau_{s}\right)}{\mu}\right]^{\omega}\vert \mathsf{S}\vert^{\left(1-\omega\right)}, \label{EQ5}
\end{align}
where notion $\tau_{s}=u\left[\operatorname{deg}\left(s\right)\right]$ traverses all nodes in graph $\mathsf{G}_{i}\left(\mathsf{S}\mid s_{j}^{i}\right)+\varepsilon$ (here $u\left(\cdot\right)$ is the unit step function). And we mark that $\omega=u\left(\sum_{s}\tau_{s}\right)$. Moreover, we pick one node that has minimum degree in the graph. Then $\mu$ measures the number of the cliques that contain this node and $\theta$ counts the cliques in the same connected component with this node.

Here we elaborate all the detailed derivations of the upper bound (\ref{EQ5}).

\subsection{Derivations of the upper bound} 
Our derivations begin with an important property discovered by Shannon and subsequent researchers \cite{shannon1956zero,lovasz1979shannon}
\begin{align}
\log\alpha\left[\mathsf{G}_{i}\left(\mathsf{S}\mid s_{j}^{i}\right)+\varepsilon\right]\leq\Theta\leq\log\lambda\left[\mathsf{G}_{i}\left(\mathsf{S}\mid s_{j}^{i}\right)+\varepsilon\right]^{-1}. \label{EQ6}
\end{align}
Here $\lambda\left(\cdot\right)$ denotes the maximum clique value
\begin{align}
\lambda\left[\mathsf{G}_{i}\left(\mathsf{S}\mid s_{j}^{i}\right)+\varepsilon\right]=\min_{\mathsf{X}}\max_{\mathsf{K}\in K}\sum_{s\in \mathsf{K}}X_{s}, \label{EQ7}
\end{align}
where $\mathsf{X}$ is an arbitrary random distribution $\mathsf{X}=\lbrace X_{s}\mid s\in \mathsf{S}\rbrace$, and $K$ denotes the clique set of confusion graph $\mathsf{G}_{i}\left(\mathsf{S}\mid s_{j}^{i}\right)+\varepsilon$ (a clique $\mathsf{K}$ is a complete sub-graph).

Although it is non-trivial to calculate $\Theta$ directly, we can still obtain a bound of $\Theta$ based on inequality (\ref{EQ6}). We suggest that the maximum clique value $\lambda$ satisfies
\begin{align}
\lambda\geq\left[\frac{\mu}{\theta+\theta\sum_{s}\left(1-\tau_{s}\right)}\right]^{\omega}\vert \mathsf{S}\vert^{-\left(1-\omega\right)}. \label{EQ8}
\end{align}
One can immediately realize that the upper bound of $\Theta$ in (\ref{EQ5}) is derived from the combination of (\ref{EQ6}) and (\ref{EQ8}). Therefore, the validity of (\ref{EQ5}) can be ensured by proving (\ref{EQ8}). 

Considering the value of $\omega$, we can subdivide (\ref{EQ8}) into two cases:
\begin{itemize}
\item The first case corresponds to $\omega=1$, or equivalently, meaning that $E\left[\mathsf{G}_{i}\left(\mathsf{S}\mid s_{j}^{i}\right)+\varepsilon\right]\neq\emptyset$. One can see that (\ref{EQ8}) can be reformulated as
\begin{align}
\lambda\geq\frac{\mu}{\theta+\theta\sum_{s}\left(1-\tau_{s}\right)}. \label{EQ9}
\end{align}
\item The second case corresponds to $\omega=0$, or equivalently, implying that $E\left[\mathsf{G}_{i}\left(\mathsf{S}\mid s_{j}^{i}\right)+\varepsilon\right]=\emptyset$. Under this condition, (\ref{EQ8}) is equivalent to
\begin{align}
\lambda\geq\mathsf{S}^{-1}. \label{EQ10}
\end{align}
\end{itemize}
Given the above analysis, let us prove (\ref{EQ9}) and (\ref{EQ10}) respectively.

\textbf{\emph{In the first case}}, we consider the dual problem of the definition of $\lambda$
\begin{align}
\lambda=\min_{\mathsf{X}}\max_{\mathsf{K}\in K}\sum_{s\in \mathsf{K}}X_{s}=\max_{\mathsf{Y}}\min_{s\in \mathsf{S}}\sum_{\mathsf{K}\ni s}Y_{\mathsf{K}}, \label{EQ11}
\end{align}
where $\mathsf{Y}$ is an arbitrary random distribution $\mathsf{Y}=\lbrace Y_{\mathsf{K}}\mid \mathsf{K}\in K\rbrace$. Here the value assignment of $\mathsf{Y}$ is implemented as following
\begin{itemize}
\item Assuming that $\mathsf{G}_{i}\left(\mathsf{S}\mid s_{j}^{i}\right)+\varepsilon$ has $m$ connected components in total, we assign $m^{-1}$ to each connected component in $\mathsf{G}_{i}\left(\mathsf{S}\mid s_{j}^{i}\right)$. The number of connected components can be worked out by $m=1+\sum_{s}\left(1-\tau_{s}\right)$;
\item In a connected component containing $k$ cliques, we assign each clique as $\left(km\right)^{-1}$.
\end{itemize} 
Based on the value assignment described above, it is easy to know
\begin{align}
\lambda\geq\min_{s\in \mathsf{S}}\sum_{\mathsf{K}\ni s}Y_{\mathsf{K}}=\theta^{-1}\mu m^{-1},\label{EQ12}
\end{align}
which can be reorganized as
\begin{equation}
\lambda\geq\mu\left[\theta+\theta\sum_{s}\left(1-\tau_{s}\right)\right]^{-1}.\label{EQ13}
\end{equation}
Thus the validity of (\ref{EQ8}) under this condition is demonstrated.

\textbf{\emph{In the second case}}, there exists no edge in $\mathsf{G}_{i}\left(\mathsf{S}\mid s_{j}^{i}\right)+\varepsilon$. Therefore, the maximum clique in this graph is each individual node itself. Following the value assignment method we use above, one can see
\begin{align}
&\lambda\leq \max_{\mathsf{K}\in K}\sum_{s\in \mathsf{K}}X_{s}=\theta^{-1}\mu m^{-1},\label{EQ14} \\
&\lambda\geq\min_{s\in \mathsf{S}}\sum_{\mathsf{K}\ni s}Y_{\mathsf{K}}=\theta^{-1}\mu m^{-1},\label{EQ15}
\end{align}
where $\mu=1$, $\theta=1$ and $m^{-1}=\vert\mathsf{S}\vert^{-1}$. Thus, (\ref{EQ8}) is correct in this case.

In summary, we can safely derive the upper bound of $\Theta$ in (\ref{EQ5}) based on (\ref{EQ6}) and (\ref{EQ8}). 

\subsection{Optimality of the upper bound} 
Here we further prove that (\ref{EQ5}) is the supremum of $\Theta$ during information diffusion if $\mathsf{G}_{i}\left(\mathsf{S}\mid s_{j}^{i}\right)+\varepsilon$ is not a complete graph. In the opposite case, $\Theta=0$ can be directly obtained, making the upper bound estimation unnecessary.

If $\mathsf{G}_{i}\left(\mathsf{S}\mid s_{j}^{i}\right)+\varepsilon$ is not a complete graph, then we know $\sum_{s}\tau_{s}<\vert\mathsf{S}\vert$, which can be subdivided into two cases
\begin{itemize}
\item $\sum_{s}\tau_{s}=0$, meaning that there is no edge in the graph. Under this condition, the biggest independent set in $\mathsf{G}_{i}\left(\mathsf{S}\mid s_{j}^{i}\right)+\varepsilon$ is itself, implying that $\alpha=\vert\mathsf{S}\vert$. Based on (\ref{EQ5}), we know 
\begin{align}
\alpha=\lambda^{-1}. \label{EQ16}
\end{align}
Combined with (\ref{EQ6}), (\ref{EQ16}) implies 
\begin{align}
\nexists \delta>0, \Theta+\delta\leq\log\lambda^{-1}. \label{EQ17}
\end{align}
Thus the upper bound in (\ref{EQ5}) is a supremum in this case.
\item $0<\sum_{s}\tau_{s}<\vert\mathsf{S}\vert$, meaning that there is at least one isolated node in the graph. Being isolated, the node has the smallest degree $0$. We know $\mu=1$ and $\theta=1$ under this condition. Then, (\ref{EQ5}) can be written as
\begin{align}
\lambda^{-1}\leq 1+\sum_{s}\left(1-\tau_{s}\right), \label{EQ18}
\end{align}
where $\sum_{s}\left(1-\tau_{s}\right)$ measures the number of isolated nodes, and $1+\sum_{s}\left(1-\tau_{s}\right)$ counts the number of connected components. 

Because the maximum independent set contains only one node from each connected component, we can know $\alpha=\lambda^{-1}$. Following (\ref{EQ17}), we can prove that the upper bound is actually a supremum in this case.
\end{itemize}

To conclude, we know that when $\mathsf{G}_{i}\left(\mathsf{S}\mid s_{j}^{i}\right)+\varepsilon$ is not a complete graph, the upper bound is a supremum.

\begin{figure*}[!t]
\centering  
\includegraphics[width=2\columnwidth]{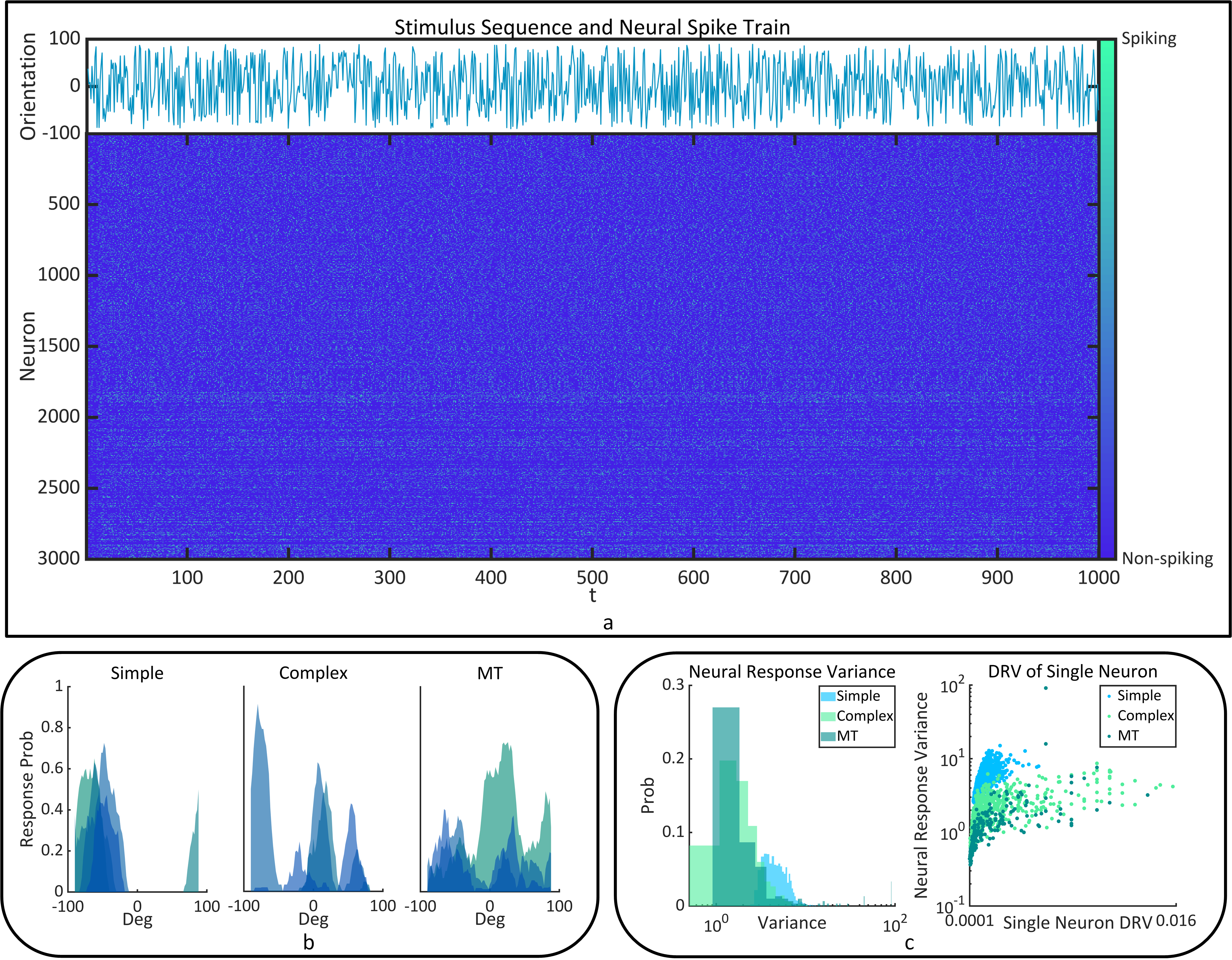}
\caption{\textbf{Characteristics of neural activities.} \textbf{a}, The randomized stimulus sequence and the corresponding stimulus-triggered neural activities are shown. \textbf{b}, We illustrate 5 examples of $\mathsf{P}\left(\text{Response}\mid\text{Stimulus}\right)$ for each kind of neurons, where one can qualitatively see the broadening process of neural selectivity from simple to MT neurons. \textbf{c}, We visualize the probability distributions of the variance of normalized neural response rates (left). One can find a positive correlation between this variance and the single neuron DRV. Similar phenomenon has be seen between the variance and the DRV in our main text.}
\end{figure*}

\section{Biological system experiment}\label{SECG}
As described in the main text, we attempt to explore the neural pathway from the primary visual cortex (V1) to the middle temporal visual cortex (MT) in the brain. The phenomenon of interest during information diffusion is that the neural selectivity (a kind of information selectivity that governs neural activity profile) changes from the selectivity of the
velocity component orthogonal to the preferred
spatial orientation (simple and complex neurons in V1 \cite{adelson1982phenomenal}) to the selectivity of velocity entirety (MT neurons \cite{rust2006mt}). Previous studies have demonstrated that this variation accounts for the subdivided and staged neural representation of motion information \cite{rust2006mt}.

 Computationally, Simoncelli and Heeger simulate the above process in a layered neural cluster model \cite{simoncelli1998model,simoncelli1996testing,simoncelli1994velocity}.  
The cluster begins with a layer to compute the weighted sum of inputs in the linear receptive
field of every simple neuron. Then, each complex neuron in the second layer responds to the weighted sum of the simple neuron afferents that are distributed within a specific spatial area and have the same orientation and phase. In the third layer, every MT neuron is driven by multiple complex neurons whose preferred orientations are consistent with the desired velocity. This model is generalized and experimentally-validated in a follow-up study \cite{rust2006mt}.

 \begin{figure*}[!t]
\centering  
\includegraphics[width=2\columnwidth]{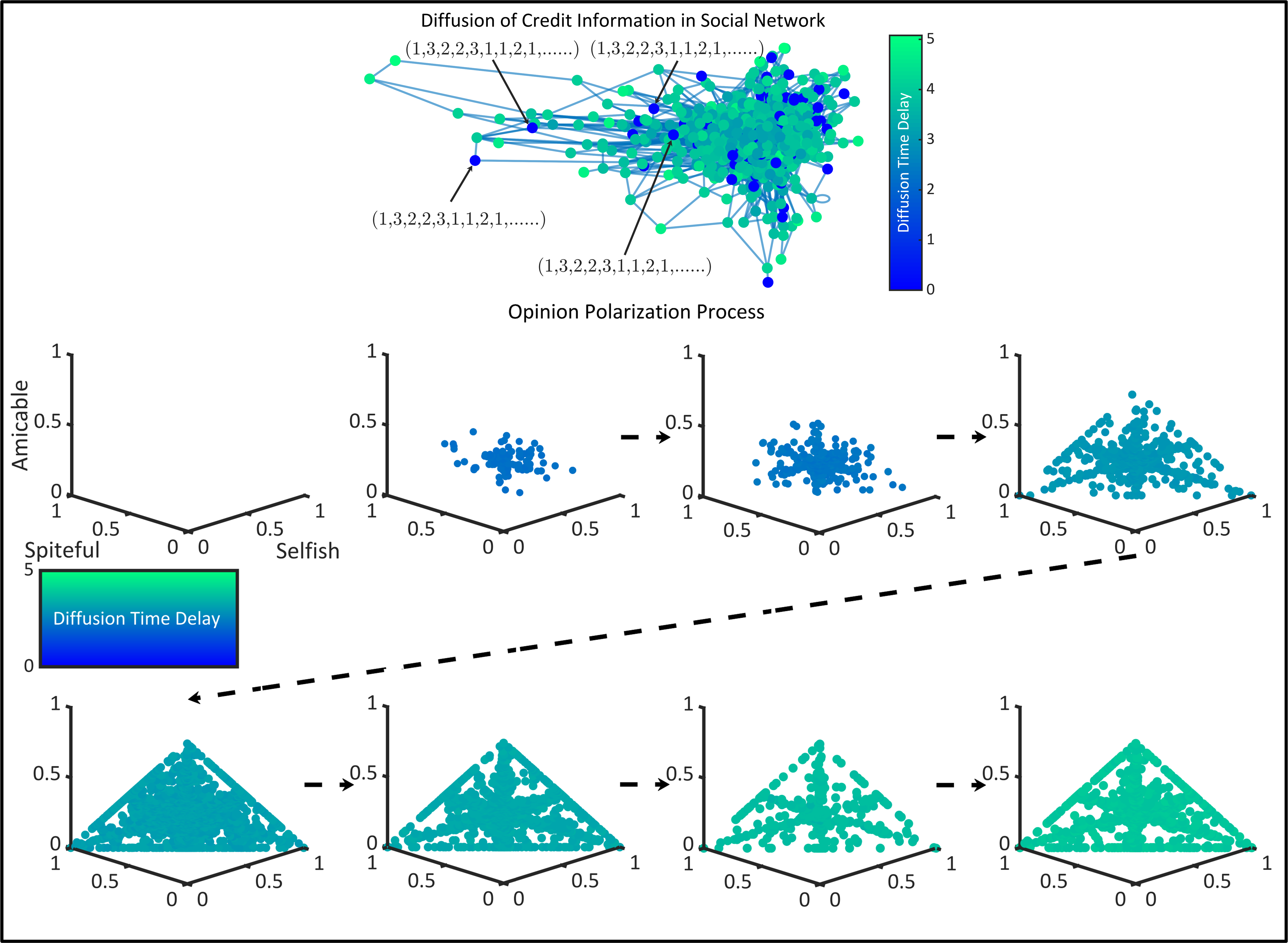}
\caption{\textbf{Opinion polarization.} The opinion polarization process during the credit information diffusion in a social network. Here all data sets obtained in 10 times of experiments are included.}
\end{figure*}

To explore how the modeled phenomenon naturally emerges from neural collective dynamics during information diffusion, we randomize a tripartite neural cluster that is not strictly layered: 
\begin{itemize}
\item The network topology of neural cluster is generated as following
\begin{itemize}
\item We randomize three complex networks, corresponding to simple, complex, and MT neuron sub-clusters. Each network is a connected graph. The number of neurons is set as $10000$, and the ratio between these three types of neurons approximates $6:3:1$. 
\item We randomly generate edges (synaptic connections) between the simple neuron sub-cluster and the complex neuron sub-cluster. About $x\%$ of complex neurons feature such connections, and each of them connects with $y$ simple neurons (here $x\approx 80$ and $y\in\left[15,50\right]$). Similar settings are applied to generate edges between the complex neuron sub-cluster and the MT neuron sub-cluster as well. The probability for a MT neuron to connect with a simple neuron directly is set as $\mathsf{P}\leq 10^{-3}$.
\end{itemize}
\item The neural activity profiles of three types of neurons are set as
\begin{itemize}
\item The information selectivity of each simple neuron is randomized following neural realistic settings (e.g., receptive filed and tuning curve \cite{ringach2002orientation,ringach2002suppression}). Specifically, the selectivity of each simple neuron is described by triangular orientation tuning curve \cite{ringach2002orientation}
\begin{align}
R\left(s\right)=u\left(B-\vert s-\hat{s}\vert\right)r_{max}\left(1-\frac{\vert s-\hat{s}\vert}{B}\right)+r_{min}, \label{EQ19}
\end{align}
where $u\left(\cdot\right)$ denotes the unit step function, $s\in\left[-90,90\right]$ denotes the stimulus orientation, notion $\hat{s}$ is the preferred orientation, parameter $B\in\left(0,180\right)$ measures the band width, and $r_{max},\;r_{min}\in\left[0,1\right]$ respectively stand for the maximum and minimum response rates \cite{ringach2002orientation}. As for complex and MT neurons, there is no preset limitation for their activity profiles. 
\item We characterize stimulus-triggered neural activities utilizing a non-homogeneous stochastic neural network \cite{tian2021characteristics,tian2021bridging}, where we treat simple neurons as input neurons in the network to drive the whole neural cluster. This framework can generate variable neural activities governed by both neural selectivity and network dynamics (see \textbf{Fig. 10a} and \textbf{Fig. 10b}). One can find a systematic definition of it in \cite{tian2021characteristics,tian2021bridging}.
\end{itemize}
\end{itemize}

Given the stimulus-triggered neural activities of our neural cluster, there are several parameters to calculate:
\begin{itemize}
\item We count the response rate $r_{i}\left(s\right)$ of each neuron $N_{i}$ to every stimulus $s$. Then we define the normalized neural response rate as 
\begin{align}
\widehat{r}_{i}\left(s\right)=r_{i}\left(s\right)\mathbb{E}_{s}\left(r_{i}\left(s\right)\right)^{-1} \label{EQ20}
\end{align}
(here the expectation value acts as the normalization reference) and calculate its variance $\operatorname*{Var}_{s}\left[\widehat{r}_{i}\left(s\right)\right]$. This variance principally measures the variability of neural responses to different stimuli. A larger response variability implies a narrower neural selectivity (see \textbf{Fig. 10c});
\item We estimate the neural response conditional probability as $\mathsf{P}_{i}\left(\text{Spike}\mid s\right)\approx r_{i}\left(s\right)f^{-1}\left(s\right)$ (here $f\left(\cdot\right)$ denotes frequency), based on which we measure $\mathsf{P}_{i}\left(\text{Spike},s\right)=\mathsf{P}_{i}\left(\text{Spike}\mid s\right)\mathsf{P}\left(s\right)$ and further define the determinability rate of neuron $N_{i}$ to stimulus $s$ as 
\begin{align}
\eta_{i}\left(s\right)=1-\mathsf{P}_{i}\left(\text{Spike},s\right)\mathbb{E}_{s}\left[\mathsf{P}_{i}\left(\text{Spike},s\right)\right]^{-1}. \label{EQ21}
\end{align}
By calculating the variance $\operatorname*{Var}_{s}\left[\eta_{i}\left(s\right)\right]$, we can quantify the capacity of neuron $N_{i}$ to resist information distortion (referred to as single neuron DRV, see \textbf{Fig. 10c}). Here we do not concentrate on $\operatorname*{Var}_{s}\left[\eta_{i}\left(s\right)\right]$ since its connection to the neural response variability $\operatorname*{Var}_{s}\left[\widehat{r}_{i}\left(s\right)\right]$ is trivial (one can immediately find the positive correlation between them). To implement non-trivial analyses, we define the determinability rate variance of neuron $N_{i}$ as $\mathbb{E}_{N_{j}}\{\operatorname*{Var}_{s}\left[\eta_{i}\left(s\right)\right]\}$, where each $N_{j}$ is a neuron located at specific diffusion paths from simple neurons to neuron $N_{i}$ (referred to as DRV). The non-triviality of this parameter lies in that it quantifies the capacity of the diffused information to resist distortion before neuron $N_{i}$ receives the information. Therefore, there can be potential causal connections between the determinability rate variance and the neural selectivity of neuron $N_{i}$ (see \textbf{Fig. 4b} in our main text).  
\end{itemize}

\section{Social system experiment}\label{SECH}
As described in the main text, we attempt to explore whether information diffusion characteristics alone are sufficient to polarize opinions in multi-agent interactions. 

Without loss of generality, we implement the analysis based on the opinion concerning credit. In real financial, marketing, and other social activities, agents may do selfish (e.g., lie or cheat) or even spiteful (e.g., break rules for non-interest reasons) behaviors to make profits or harm others \cite{hamilton1970selfish,gardner2004spite}. These costly behaviors eventually lead to the damage on credit \cite{fulker2021spite,weigelt1988reputation,kreps1990game}. A widespread phenomenon concerning credit damage is the emergence of extreme views towards the credit of an agent. The opinions on credit tend to be polarized when credit information diffuses \cite{dandekar2013biased}. Although the agent occasionally does selfish or spiteful behaviors, its credit in others' view may still approach extraordinarily high or extremely low. This phenomenon may be caused by both psychological and physical factors \cite{dandekar2013biased,lord1979biased}.

In our computational experiment, agent $i$ does selfish or spiteful behaviors with probability of $p^{\prime}$ or $p^{\prime\prime}$ in a $l$-run game, respectively. We implement the experiment in a complex network of $2000$ agents $10$ times. Each time we randomize the credit information as a string of length $l=300$, consisting of indexes $1$ (selfish), $2$ (spiteful), and $3$ (amicable). The proportion of each index in the information content represents the probability of corresponding behaviors, determining whether agent $i$ will be treated as selfish, spiteful, or amicable. The credit information first arrives at $m$ randomly selected agents ($m\approx 200$) and then diffuses to other agents (see \textbf{Fig. 11}). 

In our main text, the probability for agent $i$ to do selfish or spiteful behaviors in the game is driven farther from $p^{\prime}$ or $p^{\prime\prime}$ during information diffusion. It gradually approaches $0$ or $1$, suggesting that the opinion on the credit of agent $i$ is polarized (see \textbf{Fig. 11}). To quantify the polarization degree of opinions, we define the opinion monotonousness as
\begin{align}
\chi=\max_{i}\mathsf{P}_{i}-\operatorname*{secmax}_{i}\mathsf{P}_{i}, \label{EQ22}
\end{align}
 where $\mathsf{P}_{i}$ denotes the proportion of each index $i$ in information content (here $i\in\{1,2,3\}$), and operator $\operatorname*{secmax}$ measures the second largest value. One can see that (\ref{EQ22}) quantifies the degree for a certain index $i$ to surpass the other two indexes in proportion.
 
Moreover, we measure the proportion of extreme opinions among all opinions. Specifically, we treat an opinion as extreme if it characterizes the agent as a selfish, spiteful, or amicable person with a probability greater than $0.8$ (in other words, $\mathsf{P}_{i}>0.8$).

\nocite{1}
\bibliography{aipsamp}

\end{document}